\documentclass[useAMS,twocolumn,usenatbib,fleqn]{mn2e} \voffset=-0.8in

\usepackage{etoolbox} \usepackage{ragged2e} \usepackage{xcolor} \usepackage{fontawesome}

\usepackage[hidelinks,draft]{hyperref} \usepackage{graphicx} \usepackage{times} \usepackage{amsmath}

\newcommand{\tpara}{\delta\phi_{\parallel}} 

\newcommand{\tparaL}{\langle\tpara\rangle}

\newcommand{\rapo}{R_{\rm apo}} \newcommand{\vapo}{V_{\rm apo}}  \newcommand{\rperi}{R_{\rm peri}}  \newcommand{\rturn}{R_{\rm turn}} \newcommand{\vturn}{V_{\rm turn}} \newcommand{\vvir}{V_{\rm vir}}

\def\kms{ {\rm km~s\textsuperscript{-1}}}

\usepackage{pifont}

\defcitealias{petersen16a}{PWK16} \defcitealias{petersen18b}{Paper II} \defcitealias{petersen18c}{Paper III}

\usepackage{natbib} 

\begin{document}

\title[Commensurabilities in Disc Galaxies] {Using Commensurabilities and Orbit Structure to Understand Barred Galaxy Evolution}

\author[Petersen, Weinberg, \& Katz] {Michael~S.~Petersen,$^{1,2}$\thanks{michael.petersen@roe.ac.uk} Martin~D.~Weinberg,$^2$ Neal~Katz$^2$\\ $^1$Institute for Astronomy, University of Edinburgh, Royal Observatory, Blackford Hill, Edinburgh EH9 3HJ, UK \\ $^2$Department of Astronomy, University of Massachusetts at Amherst, 710 N. Pleasant St., Amherst, MA 01003}

\maketitle

\begin{abstract} We interpret simulations of secularly-evolving disc galaxies through orbit morphology. Using a new algorithm that measures the volume of orbits in real space using a tessellation, we rapidly isolate commensurate (resonant) orbits. We identify phase-space regions occupied by different orbital families. Compared to spectral methods, the tessellation algorithm can identify resonant orbits within a few dynamical periods, crucial for understanding an evolving galaxy model. The flexible methodology accepts arbitrary potentials, enabling detailed descriptions of the orbital families. We apply the machinery to four different potential models, including two barred models, and fully characterise the orbital membership. We identify key differences in the content of orbit families, emphasising the presence of orbit families indicative of the bar evolutionary state and the shape of the dark matter halo. We use the characterisation of orbits to investigate the shortcomings of analytic and self-consistent studies, comparing our findings to the evolutionary epochs in self-consistent barred galaxy simulations. Using insight from our orbit analysis, we present a new observational metric that uses spatial and kinematic information from integral field spectrometers that may reveal signatures of commensurabilities and allow for a differentiation between models. \end{abstract}


\section{Introduction} \label{sec:introduction}

Empirically, a typical axisymmetric disc is dominated by orbits that appear to be regular rosettes of varying eccentricities determined by the radial energy. Such regular orbits have constants of motion and are considered integrable, and can be represented by a combination of three fundamental frequencies. As stated by the Jeans theorem \citep{jeans15} for an axisymmetric system, the distribution function is a function of isolating integrals of motion. For example, in a razor-thin disc, these would be energy $E$ and the angular momentum about the galaxy's symmetry axis $L_z$. This principle has been used in a number of analytic studies over the past century \citep[see][]{binney08}, including recent advancements \citep{binney16}. Additionally, the Jeans equations \citep{jeans22} have been used to perform assessments of the content of orbital families in a model -- the {\sl orbital structure} -- of real galaxies under the assumption that the galaxies can be described by the classical integrals of motion: $E$, $L_z$, and a third integral commonly referred to as $I_3$ \citep{nagai76, satoh80, binney90, vandermarel90}.

Unfortunately, the assumption that galaxies are semi-isotropic rapidly breaks down for realistic galaxies and dynamical models. While typical rosette orbits in an axisymmetric system fill an area of the disc after many orbital periods, non-axisymmetries in the system may create new families of commensurate (or resonant) orbits\footnote{In addition to commensurabilities that exist in the axisymmetric disc-halo system.}. These commensurate orbits are defined by the equation \begin{equation} m\Omega_p=l_1\Omega_r + l_2\Omega_\phi+l_3\Omega_z \label{eq:resonances} \end{equation} where $\Omega_{r,\phi,z}$ are the polar coordinate frequencies of a given orbit and $\Omega_p$ is some pattern frequency, e.g. the frequency of a bar or spiral arms. Commensurate orbits are closed curves and have formally zero volume, in the sense of the configuration-space volume filled by the orbit, when pictured in a frame rotating with the pattern speed $\Omega_p$. One may consider the commensurate orbits as the backbone around which secular evolution occurs -- where $E$ or $L_z$ can be gained or lost, for example.

Even in the case of relatively simple potentials, such as an exponential stellar disc embedded in a spherical dark matter halo, finding the distribution function, fundamental frequencies, and/or commensurate orbit families analytically can rapidly become intractable, especially in the presence of a modest to strong non-axisymmetric perturbation. Any known method for producing analytic potentials is insufficient for characterising the real universe. Few axisymmetric potentials that resemble real galaxies can be described via separable potentials \citep{dezeeuw85}. Further, the inclusion of non-axisymmetric features, such as a bar, can render the potential calculation intractable\footnote{For simple analytic bar potential expressions, extensions of analytic studies are able to make some progress \protect{\citep[e.g.][]{binney18}}.}. Simply changing the halo model from a central cusp to a central core is known to alter the families of bar orbits present near the centre of the galaxy \citep{merritt99}. Thus, it is difficult to constrain the orbital structure of realistic, evolving galaxies. The lack of techniques in the literature for determining orbital families in evolving potentials applicable to realistic galaxies (e.g. non-axisymmetries) motivates finding new methodologies that determine the orbital content of a disc galaxy. We present techniques suitable for studying orbits in evolving potentials in this paper.

Analytic and idealised numerical studies of potentials representing barred galaxies show a basic resonant structure that underpins the bar represented by the commensurate $x_1$ orbit, which arises from the inner Lindblad resonance \citep[ILR, where $2\Omega_p=-\Omega_r+2\Omega_\phi$; ][]{contopoulos80,contopoulos89,athanassoula92,skokos02}. However, small adjustments to the mass of the model bar can admit new commensurate subfamilies of $x_1$ orbits, necessitating a model-by-model (or galaxy-by-galaxy) orbital census. Such subfamilies have been extensively documented in the literature \cite{contopoulos89}; we will discuss and show subfamilies below. For simulated galaxies, taking the measure of the orbital structure means being able to (nearly) instantaneously identify the family of a given orbit, as the orbit family may change over a handful of dynamical times. Bifurcations of prominent orbit families result from alterations to the potential shape, giving rise to families such as the 1/1 (sometimes stylised 1:1) orbit, a family which results from the bifurcation of the $x_1$ family \citep{contopoulos83, papayannopoulos83, martinet84, petrou86}. For consistency, we refer to this orbit throughout this work as an $x_{1b}$ orbit, denoting that the family is a bifurcation of the standard $x_1$ orbit.

Finding commensurabilities in realistic potentials has proven particularly difficult in an analytic framework \citep{binney08}. Even more difficult is identifying the act of `trapping', or capture into resonant orbits, which cannot occur in fixed gravitational potentials. Recently, techniques to describe orbits observed in self-consistent simulations such as those drawn from $N$-body simulations, i.e., those that are allowed to evolve with gravitational responses, have been used to find the rate at which particles change orbit families in response to the changing phase-space structure \citep[][hereafter PWK16]{petersen16a}.

\begin{table*} \label{tab:models} \begin{tabular}{lccccccc} \hline Potential Number & Simulation Name, Time & Potential Name & Scalelength &Disc Mass & M$_{\rm halo}(<R_{\rm d})$ & $r_{M_h = M_d}$ & Pattern Speed\\ & & &$R_{\rm d}$ [$R_{\rm vir}$] & [$M_{\rm vir}$] &[$M_{\rm vir}$] & [$R_{\rm vir}$] & $\Omega_p$ [rad/$T_{\rm vir}$]\\ \hline I &Cusp Simulation $T=0$ &Initial Cusp & 0.01 & 0.025 & 0.0050 & 0.0167 & 90 \\ II & Cusp Simulation $T=2$ & Barred Cusp & 0.01 & 0.025 & 0.0050 & 0.0172 & 37.5 \\ III & Core Simulation $T=0$& Initial Core & 0.01 & 0.025 & 0.0022 & 0.0317 & 70 \\ IV & Core Simulation $T=2$ & Barred Core & 0.01 & 0.025 & 0.0024 & 0.0282 & 55.6 \\ \hline \end{tabular} \caption{Potential models used in the detailed fixed potential study.} \end{table*}

Despite the effort placed into studying both analytic potentials and self-consistent simulations, a vast gulf of understanding exists between analytic and self-consistent potentials. Linear or weakly non-linear galaxy dynamics can only be extended so far: at some point, the distortions become so strong that it is insufficient to consider perturbations to the system and one must treat the entire system in a self-consistent manner. However, fully self-consistent simulations are encumbered by the many parameters necessary to describe a galaxy, all of which are difficult to control when designing self-consistent model galaxies. Designing model galaxies that match observations of real galaxies, including the Milky Way (MW), is a challenging process. Fixed-potential orbit analysis is often used as a bridge between analytic and self-consistent work.

Within this framework, many previous studies have used frequency analysis to characterise the properties of orbits, describing orbits by their frequencies in various independent dimensions. The result is a partitioning of orbits into families that reside on integer relations between frequencies, using equation~(\ref{eq:resonances}). Attempting frequency analysis on an ensemble of real orbits is a natural extension, via either spectral methods \citep{binney82, binney84} or frequency mapping \citep{laskar93, valluri12, valluri16}. However, these tools are only useful when the evolution of the system is slow, or the evolution is artificially frozen. Therefore, we develop a new methodology based on a simple and robust tessellation algorithm that permits the unambiguous determination of commensurate orbits on several orbital time scales enabling analysis in evolving simulations. We emphasise that the utility of this method extends beyond the proof-of-concept presented here. An advantage to this orbit atlas analysis is its ability to move beyond the standard methods of locating resonances based on frequencies. We are able to empirically determine the location of all closed orbits in both physical and conserved-quantity space.

In this paper, we apply this simple methodology and characterise orbital structure, with a particular emphasis on commensurate orbits, to understand barred galaxy evolution. This work presents significant upgrades to one orbit analysis tool previously published \citepalias{petersen16a}, as well as an entirely new algorithm. The goal of this project is to develop new techniques for comparing non-axisymmetric orbits between fixed potential simulations and fully self-consistent simulations. We discern distinct phases or epochs of evolution of a bar in self-consistent simulations. Along the way, we demonstrate that (1) we can efficiently dissect bar orbits into dynamically relevant populations, (2) commensurate orbit families can be efficiently found and tracked through time across different fixed-potential realisations, (3) commensurate orbits provide a useful method to analyse self-consistent simulations, and (4) one may infer the evolutionary phase of barred galaxies from this methodology.

The organisation of the paper is as follows. We describe the models we studied in the course of this work and present new techniques in Section~\ref{sec:methods}. Results from different fixed potential models are presented in Section~\ref{sec:results}. We then discuss the implications of the findings for interpreting other models in Section~\ref{subsec:selfconsistent}. We discuss the implications of the results for observational studies in Section~\ref{sec:observations}. We then use the lessons from our fixed potential analysis to interpret the evolution in the self-consistent simulations in Section~\ref{sec:selfconsistent}. We conclude and propose future steps in Section~\ref{sec:conclusion}.

\section{Methods} \label{sec:methods}

We first present the initialisation and execution of self-consistent disc and halo simulations in Section~\ref{subsec:simulations}. An overview of the improved apoapse clustering orbit classifier for closed orbit identification presented in \citetalias{petersen16a} is discussed in Section~\ref{subsec:trapping}. The potentials we use for detailed study of fixed-potential integration are described Section~\ref{subsec:fixedpotentials} and the determination of the bar position and pattern speed in Sections~\ref{subsubsec:bardetermine} and \ref{subsubsec:figure_rotation}, respectively. In Section~\ref{subsec:orbit_atlas}, we describe the creation of an representative ensemble of orbits for each model that we call an \emph{orbit atlas}, including the initial condition population (Section~\ref{subsubsec:initial_conditions}), and integration method (Section~\ref{subsubsec:integration}).

\subsection{Simulations} \label{subsec:simulations}

\begin{figure} \centering \includegraphics[width=3.4in]{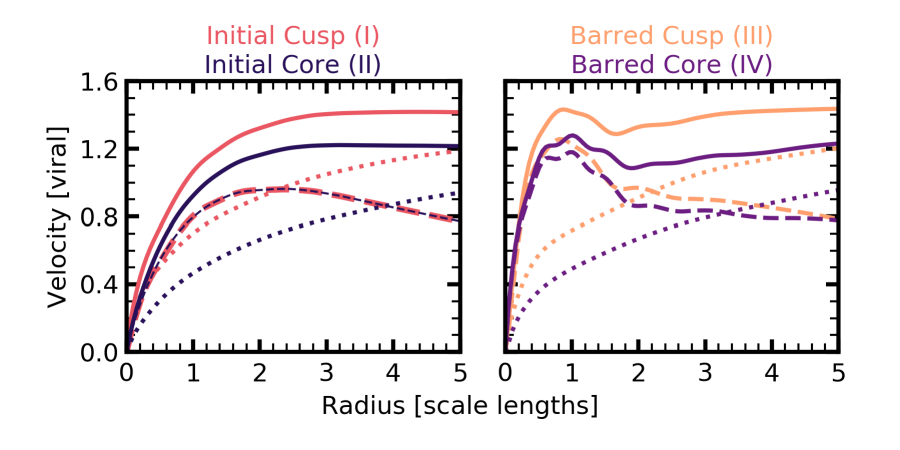} \caption{\label{fig:circular_velocity} Circular velocity curves as a function of radius, computed for the cusp and core simulations at $T=0$ and $T=2$. The left panel shows the two initial disc models ($T=0$, the initial conditions of each simulation), while the right panel shows the two barred models ($T=2$, after moderate evolution in each simulation). Both panels are colour coded as shown above the panels. The solid lines are the circular velocity at each radius computed from the monopole for the total system. The dashed (dotted) lines are the monopole-calculated circular velocity for the disc (halo) component only.} \end{figure}

We employ two galaxy simulations, loosely inspired by the MW, in this work. The simulations used here are updated slightly from the simulations presented in \citetalias{petersen16a}, including a modestly more concentrated halo and significantly longer time integration. We justify both changes at the end of this section.

\subsubsection{Initial Conditions} \label{subsubsec:initialconditions}

Both simulations feature an initially spherically-symmetric Navarro-Frank-White (NFW) dark matter halo radial profile \citep{navarro97}, which we generalise to include a core where the density $\rho_h(r)$ becomes constant with radius: \begin{equation} \rho_h(r) = \frac{\rho_0r_s^3}{\left(r+r_c\right)\left(r+r_s\right)^2} \label{eq:nfw} \end{equation} where $\rho_0$ is a normalisation set by the chosen mass, $r_s=0.04R_{\rm vir}$ is the scale radius, $R_{\rm vir}$ is the virial radius, and $r_c$ is a radius that sets the size of the core. $r_s$ is related to the concentration, $c$, of a halo by $r_s = R_{\rm vir}/c$. The halo has $c=25$, consistent with a normal distribution of halo concentrations from recent cosmological simulations \citep{fitts18,lovell18}. We include an error function truncation outside of $2R_{\rm vir}$ to give a finite mass, $\rho_{\rm halo, trunc}(r) = \rho_{\rm halo}(r)\left[\frac{1}{2}-\frac{1}{2}\left({\rm erf}\left[(r-r_{\rm trunc})/w_{\rm trunc}\right]\right)\right]$, where $r_{\rm trunc}=2R_{\rm vir}$ and $w_{\rm trunc}=0.3R_{\rm vir}$.

The normalisation of the halo is set by the choice of virial units for the simulation, such that $R_{\rm vir} = M_{\rm vir} = v_{\rm vir} = T_{\rm vir} = 1$. Scalings for the MW halo mass suggest that $R_{\rm vir} = 300~{\rm kpc}$, $M_{\rm vir} = 1.4\times 10^{12}~{\rm M_\odot}$, $v_{\rm vir} = 140~\kms$, and $T_{\rm vir} = 2~{\rm Gyr}$. The motivation behind generalising the NFW profile to include a core lies in the ambiguity of the central density of dark matter halos in observed galaxies, including the MW \citep{mcmillan17}. The pure NFW profile extracted from dark matter-only simulations is cuspy. The first two potentials have $r_c=0$ and, therefore, we refer to these as `cusp' potentials (Table~\ref{tab:models}).

We embed an exponential disc in the 'cusp' potential halo in our first 'cusp simulation'. The three-dimensional structure of the disc is given as an exponential in radius and an isothermal ${\rm sech}^2$ distribution in the vertical dimension: \begin{equation} \rho_d(r,z) = \frac{M_{\rm d}}{8\pi z_0R_d^2} e^{-r/R_d} {\rm sech}^2 (z/z_0) \label{eq:exponentialdisc} \end{equation} where $M_d=0.025M_{\rm vir}$ is the disc mass, $R_d=0.01R_{\rm vir}$ is the disc scale length, and $z_0=0.001R_{\rm vir}$ is the disc scale height, which is constant across the disc. We set $r_c=0.02R_{\rm vir}(=2R_d)$ for the second simulation, and we refer to that simulation as the `core simulation'. We tailor $\rho_0$ for the cored simulation initial condition such that the virial masses are equal to that of the cusp simulation, i.e. $M_{\rm vir, cusp} = M_{\rm vir, core} = 1$. We again embed a 0.025$M_{\rm vir}$ initially exponential disc in this halo (Table~\ref{tab:models}).

Both simulations presented here have $N_{\rm disc}=10^6$ and $N_{\rm halo}=10^7$, the number of particles in the disc and halo component, respectively. The disc particles have equal mass. We employ a `multimass' scheme for the halo to increase the number of particles in the vicinity of the disc. The procedure to generate a multimass halo is as follows.

Let $n_{\rm halo}(r)$ be the desired number density profile for halo particles and $\rho_{\rm halo}(r)$ be the desired halo mass density. We solve the Abel integral equation using a generalised Osipkov-Merritt parametrisation as a function of $E$ and $L$ \citep{binney08} to obtain the corresponding number and mass distribution functions $f_{\rm number}$ and $f_{\rm mass}$\footnote{In the particular realisations for this paper, we define the anisotropy radius $r_a=\infty$, such that finding the distribution functions reduces to the standard Eddington inversion. However, our approach does not require this choice.}. We realise a phase-space point by first selecting random variates in energy $E$ and angular momentum $L$ by the acceptance-rejection technique from the distribution $f_{\rm number}$. Then, the orientation of the orbital plane and the radial phase of the orbit are chosen by uniform variates from their respective ranges. This determines position and velocity. Finally, we set the mass of the particle, $m=(M_{\rm halo}/N_{\rm halo}) f_{\rm mass}(E, L)/f_{\rm number}(E, L)$. We choose a target number density profile for our simulation particles $n_{\rm halo}\propto r^{-\alpha}$ with $\alpha=2.5$.

The number density of particles in the inner halo, $r<0.05R_{\rm vir}(=5R_d)$, is improved by roughly a factor of 100, making the mass of the average halo particle in the vicinity of the disc equal to that of the disc particles. This is equivalent to using $10^9$ equal mass halo particles. For the basis-function expansion, this implies a higher signal to noise ratio for halo coefficients that affect the gravitational field in the vicinity of the disc.

The disc velocities are chosen by solving the Jeans' equations to second order, with asymmetric drift, in cylindrical coordinates in the combined disc--halo potential, also as in \citetalias{petersen16a}. The equations are found in \citet{binney08}. The radial velocity dispersion is set by the choice of the Toomre $Q$ parameter such that \begin{equation} \sigma_r^2(r) = \frac{3.36\Sigma(r)Q}{\kappa(r)} \label{eq:toomre} \end{equation} where $\Sigma(r)$ is the disc surface density, and the epicyclic frequency, $\kappa$, is given by \begin{equation} \kappa^2(r) = r\frac{d\Omega_c^2}{dr}+4\Omega_c^2. \end{equation} where $\Omega_c$ is the circular orbit azimuthal frequency. Our choice of $Q=0.9$ is motivated by our desire to form a bar in a short time period. While our initial conditions are unlikely to be found in the real universe, the rapid formation of a bar provides an opportunity to study the evolution of a realistic system that resembles observed galaxies (see Figure~\ref{fig:surfacedensity}).

The disc velocities in each of the three cylindrical dimensions $(R,\phi,z)$ are realised using a multivariate Gaussian distribution as \begin{equation} \begin{split} v_R &= x_1\sigma_R\\ v_\phi &= x_2\sigma_\phi + \bar{v}_{\phi}\\ v_z &= x_3\sigma_z \end{split} \end{equation} where $x_1,x_2,x_3$ are three random normally-distributed variables and the mean azimuthal velocity $\bar{v}_\phi$ is defined as \begin{equation} \bar{v}^2_\phi(r) = v_{\rm circ}^2(r) + \frac{R}{\rho_{\rm disc}(R)}\frac{d}{dR}\left[\rho_{\rm disc}(R)\sigma_R^2(R)\right] + \sigma_R^2(R) - \sigma_\phi^2(R) \end{equation} where $v_{\rm circ}$ is the circular velocity derived from the potential and the other terms on the right-hand side correspond to the asymmetric drift. We assume $\sigma^2_R=\sigma^2_\phi$ for these initial conditions.

To allow the disc and halo to reach a mutual equilibrium, we evolve the particles in a fixed disc potential, but live halo potential, until $T=2T_{\rm vir}$ and then restart the simulation with the evolved positions and velocities. In practice, this allows (1) the halo to contract along the $z$ axis in the presence of the disc, which is only accounted for as a spherical monopole in the initial spherical halo potential, and (2) any initial rings resulting from imperfect equilibrium in the disc to dissipate. These rings result from the closure of the Jeans' equations at second order, as well as a theoretical breakdown of the Jean's equations at $r<z_0$, and are $\mathcal{O}(10^{-2}\rho_{\rm disc,0})$ at all radii, where $\rho_{\rm disc,0}$ is the initial disc density. The rings phase mix within 20 local crossing times. We do not see any evidence that the results of our study are affected by the rings. We have checked that the disc density distribution remains unchanged.

As discussed in \citetalias{petersen16a}, the maximum contribution to the total circular velocity by the disc, $f_D\equiv V_{{\rm circ},\star}/V_{{\rm circ},{\rm tot}}$, for typical disc galaxies is $f_D=0.4-0.7$, with $\langle f_D\rangle=0.57$ \citep{martinsson13}. Our cusp simulation has $f_D=0.65$ and our core simulation has $f_D=0.75$. With the new simulations, we evolve until $T=4.5T_{\rm vir}$. For a MW-mass galaxy, this is equivalent to 9 Gyr. We acknowledge that it is unrealistic to expect that a galaxy will evolve in a purely secular fashion for half the age of the universe, without interactions or mass accretion. However, integrating the simulations for a substantial time allows for a full range of evolutionary states to develop as discussed below, which help to probe the dynamical mechanisms behind bar evolution in the real universe.

\subsubsection{N-body Simulation} \label{subsubsec:nbody}

To integrate not only the $N$-body model but full orbits, we require a description of the potential and force vector at all points in physical space and time. We accomplish this using a bi-orthogonal basis set of density-potential pairs.  We generate density-potential pairs using the basis function expansion (BFE) algorithm implementation {\sc exp} \citep{weinberg99}. In the BFE method \citep{cluttonbrock72, cluttonbrock73, hernquist92a}, a system of bi-orthogonal potential-density pairs are calculated and used to approximate the potential and force fields in the system.  The functions are calculated by numerically solving the Sturm-Louiville equation for eigenfunctions of the Laplacian. The full method is described in \citet{petersen20b} for both spherical and cylindrical bases. The description and study of the eigenfunctions that describe the potential and density is the focus of a companion paper \citep[][hereafter Paper III]{petersen18c}.

The BFE for halos are best represented by an expansion in spherical harmonics with radial basis functions determined by the target density profile. For a spherical halo, the lowest order radial basis function for \(l=m=0\) can be chosen to match the initial model. To capture evolution, the halo is described by $\left(l_{\rm halo}+1\right)^2\times n_{\rm halo}$ terms, where $l_{\rm halo}$ is the maximum order of spherical harmonics retained and $n_{\rm halo}$ is the maximum order of radial terms kept per $l$ order. This method will work for triaxial halos as well.  For triaxial halos, the target density can be chosen to be a close fitting spherical approximation of the triaxial model.  The equilibrium will require non-axisymmetric terms but the series will converge quickly.

The cylindrical basis representing the disc is expanded into $m_{\rm disc}$ azimuthal harmonics with $n_{\rm disc}$ radial sub-spaces. Each subspace has a potential function with corresponding force and density functions. The lowest-order disc pair matches the initial equilibrium profile of the analytic functional form given in equation~(\ref{eq:exponentialdisc}). Each of these density and potential basis functions are functions in cylindrical radius \(R\) and vertical height above the disc plane \(z\). Just like the spherical basis, the completeness of the biorthogonal expansion for the three-dimensional disc is guaranteed by the properties of the Sturm-Liouville equation. Successive terms probe finer spatial structure. We truncate the series to follow structure formation over a physically interesting range of scales, which has the added benefit of reducing small-scale noise including two body scattering. We may select different features by excluding functions where structural variations are not of interest. A covariance analysis of the coefficient evaluation shows that our truncation includes all basis terms that contribute significant signal.  That is, nearly all excluded terms in our simulations represent Poisson particle noise only.

The potential at any point in the simulation is represented by $(m_{\rm disc}+1)\times n_{\rm disc}$ coefficients for the corresponding orthogonal functions. The disc basis functions are identical between the cusp and core models. The halo basis functions are necessarily different to capture the initial density profile in the lowest-order term. We retain azimuthal and radial terms ($m_{\rm disc}$=$l_{\rm halo}=m_{\rm halo}\le6$, $n_{\rm disc}\le 12$, $n_{\rm halo} \le 20$) chosen for both the disc and halo depending upon the simulation goals. We discuss the effect on our results owing to the inclusion or exclusion of higher-order harmonics ($m=3,4,5,6$) in detail in Section~\ref{subsubsec:bcusp}. The halo has a larger number of radial ($n$) terms to probe similar scales in the disc vicinity. The disc basis is truncated at $r=0.2R_{\rm vir}$ or 20 disc scale lengths, and $|z|=0.1R_{\rm vir}$ outside of which we calculate its contribution using the monopole term only. At large distances $d$ from the disc, the contribution from terms with $m>0$ goes to zero at the rate $d^{-m}$ faster than monopole. As a result, the spatial truncation of the basis introduces only a maximum 0.3 per cent error in the integration as halo orbits cross this boundary.

{\sc exp} allows for an easy calculation of the potential from both the initial galaxy mass distribution as well as the evolved galaxy mass distribution.  The key limitation of the BFE method lies in the loss of flexibility owing to the truncation of the expansion; large deviations from the equilibrium disc or halo will not be well represented. Although the basis is formally complete, our truncated version limits the variations that can be accurately reconstructed. Despite this, basis functions can be a powerful tool to gain physical insight; analogous to traditional Fourier analysis, a BFE identifies spatial scales and locations responsible for the model evolution.

\begin{figure} \centering \includegraphics[width=3.2in]{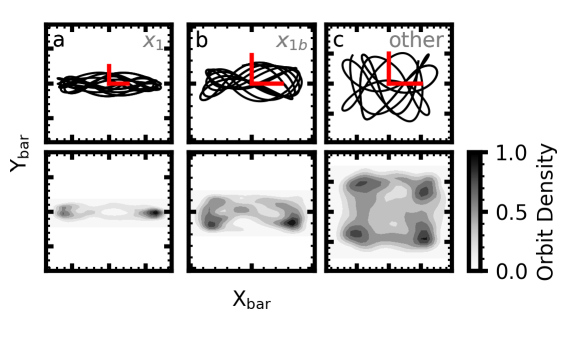} \caption[Example orbits associated with the bar drawn from the self-consistent cusp simulation.]{\label{fig:example_bar} Three primary self-consistent bar orbit families classified from the cusp simulation near $T=2$. The upper panels are the trajectories, while the lower panels are the time-integrated densities, or `orbit density' (i.e. showing where the trajectory moves faster or slower such that an orbit resides at a position for longer). The orbits are organised from largest radial extent to smallest, with the red bars indicating 0.5$R_d$ in each panel. From left to right: (a) A standard $x_1$ orbit. (b) A bifurcated $x_{1b}$ orbit. (c) An `other' bar orbit, in this case, a nearly 4:2 orbit. All orbits are plotted in the frame rotating with the bar. } \end{figure}

\subsection{Computing Trapping} \label{subsec:trapping}

We have developed several improvements to the clustering method of \citetalias{petersen16a} that enables one to determine the membership in different orbital subpopulations beyond the bar-supporting orbits during the simulations. For example, consider a periodic bar-supporting $x_1$ orbit.  The angle of the apoapses will tightly cluster at 0 and 180 degrees from the bar major axis.  The \citetalias{petersen16a} method used a one-dimensional clustering algorithm to identify this bunching of apoapse position angles or individual orbits.  From their distribution, one can identify various trapped orbit families. No conclusions from \citetalias{petersen16a} change as a result of this upgrade; rather, the improvements provide a more detailed classification. The more sophisticated algorithm builds upon the same apoapse-clustering technique, but uses additional diagnostics related to the distribution of apoapses within the $k$ clusters to determine membership in orbit families. The details of the classification procedure are described in Appendix~\ref{appendix:orbitclassification}. In this section we give a qualitative overview and discuss the theoretical motivation behind our cluster-based orbit classification.

The orbits that make up a galaxy model are both a reflection of, and support, the potential of the galaxy. The pioneering work of \citet{contopoulos80} presented a census of bar-supporting orbits, including the principal $x_{1}$ family. Called the `backbone' of the bar, $x_1$ orbits exist at various energies set by the shape of the potential. However, determining family membership in self-consistent models has been challenging. The concept of the trapping of orbits into reinforcing structures in the potential is a dynamically complex, but straightforward, process under idealised conditions. In the case of perturbation theory, one may compute a capture criterion or trapping rate \cite[e.g.][]{contopoulos78,henrard82,binney08,daniel15}, i.e., the probability that a star on a certain orbit joins a particular resonance parented by some closed commensurate orbit for which the potential may be specified.

In a self-consistent evolving galaxy, the process and probability of being captured into a resonance--and even the location of the resonance itself--is difficult to ascertain. Several techniques have focused on the use of `frozen' potentials, as follows. First, a model is evolved self-consistently up to some time. Then the potential is frozen and orbits are then integrated in the fixed potential to determine the orbital structure. We use a hybrid approach where we simultaneously analyse frozen potentials and self-consistent simulations. With input from analytic orbit family descriptions, we hope to dissect our models using the $k$-means methodology at every recorded timestep ($h=0.002 T_{\rm vir}$) to determine the constituent orbits while the simulation undergoes self-consistent evolution. We call the identification of orbit families during self-consistent evolution `in vivo' classification. In practice, this means selecting some finite time window of the orbit's evolution, typically 1-2 bar periods, in which we determine membership in an orbital family. The $k$-means classifier is largely insensitive to variations in family membership on time scales smaller than the median $x_1$ azimuthal period.

The mass that supports the bar feature is a fundamental quantity in a barred galaxy model. However, determining the trapped mass is not an easy task, as we must empirically find parameters for determining trapped orbits in vivo, and both systematic and random errors cause uncertainty. Despite this, our apoapse-clustering technique efficiently locates and identifies orbits that are members of the bar. The required time resolution is on the order of a handful of turning points per orbit.  Other classifiers rely on an instantaneous spatial or kinematic determination of the disc galaxy structure. The strength of our apoapse-clustering method is that it depends only on the positions of the turning points relative to the bar angle. This makes the methodology (1) fast and (2) independent of detailed simulation processing. The closest analogue to our procedure found in the literature is that of \cite{molloy15}, who used rotating frames to more accurately calculate the epicyclic frequency. However, their procedure is only robust for orbits that are not changing their family over multiple dynamical times. Our method is robust to orbits that are only trapped for one or two dynamical times.

We classify three primary types of bar orbit, with prototypical orbits for each shown in Figure~\ref{fig:example_bar}: \begin{enumerate} \item $x_1$ orbits, the standard bar-supporting orbit (panel a of Figure~\ref{fig:example_bar}). \item $x_{1b}$ orbits, a subfamily resulting from a bifurcation of the $x_1$ family that are often referred to in the literature as 1/1 orbits (panel b of Figure~\ref{fig:example_bar})\footnote{The so-called 1/1 orbits are a bifurcation of the $x_1$ orbit family owing to a transverse perturbation with the same frequency as the orbital radial frequency (hence 1/1, the ratio of the radial frequency of the orbit to that of the corresponding $x_1$ orbit), making a new region of phase-space become energetically favourable. In our case, this transverse perturbation is a combination of the $m>2$ components in the bar. See the studies of \protect{\citet{contopoulos83}}, \protect{\citet{papayannopoulos83}}, \protect{\citet{martinet84}}, \protect{\citet{sparke87}}.}. \item `Other' bar-supporting orbits that are coherently aligned with the bar potential but are not part of the $x_1$ family, generally demonstrating higher-order behaviour (panel c of Figure~\ref{fig:example_bar}). \end{enumerate} The orbits in Figure~\ref{fig:example_bar} are drawn from the cusp simulation as having been trapped into their respective families at $T=2$. Each orbit has the time series from the cusp simulation $T=1.8-2.2$ plotted in the upper row, with the time-averaged orbit density shown in the bottom row. We refer to the grey scale in the lower row as the `orbit density', that is, the normalised relative probability of finding an orbit at a given location in the trajectory. The time-averaged orbit density is computed using an intentionally-broadened kernel that more clearly displays the densities. We have plotted only the first five radial periods in the upper row of panels for clarity in following the trajectories. In these examples, as in most cases drawn from self-consistent simulations, the true nature of the orbit is difficult to determine from the trajectory, but becomes apparent from the time-integrated location, motivating our inclusion of the time-integrated location, or `orbit density' in space, throughout this work.

\subsection{Fixed Potentials} \label{subsec:fixedpotentials}

We select four example snapshots where we fully decompose and describe the orbit structure, and apply the general results to the evolution of barred systems in later sections.

\subsubsection{Potential Selection} \label{subsubsec:selection}

From each of the cusp and core simulations, we compute the fixed potential at two times, $T=0 T_{\rm vir}$ and $T = 2 T_{\rm vir}$, in which we will characterise the orbital structure. At each time, we compute the coefficients for the basis functions used for integration by {\sc exp}. Each particle makes a contribution to the the tabulated basis functions. We calculate the potential for the entire ensemble by accumulating the contribution from all particles in the system, resulting in coefficients that serve as the weights for the different functions. The coefficients for the initial models are calculated from the initial distribution ($T=0 T_{\rm vir}$), and are dominated by the lowest order term by construction. The coefficients for the barred models ($T=2 T_{\rm vir}$) are calculated from the self-consistent evolution of the systems. The first potential is the initial exponential disc embedded in the spherical NFW cusp halo, Potential I (Initial Cusp). We also self-consistently evolve the exponential disc to a time after a bar has formed to $T=2 T_{\rm vir}$, Potential II (Barred Cusp). Similarly, we choose the analogous time points for the cored simulation, Potential III (Initial Core), and Potential IV (Barred Core). The bar in the core self-consistent model is also still slowing and evolving, including active lengthening at the time we selected. The evolution of the core simulation is discussed in Section~\ref{subsec:coresimulation}.

Figure~\ref{fig:circular_velocity} shows the circular velocity calculated from the monopole contribution as a function of radius (solid lines). The four potentials are colour coded as indicated in the figure. In both panels, we decompose the total circular velocity into contributions from the halo (dotted lines) and disc (dashed lines). As the initial discs are the same between the cusp and core simulations, differences in the total circular velocity are caused by the halo. The halo models remain largely unchanged between the Initial and Barred version of the models, with modest ($<$10 per cent) changes to the enclosed mass between the initial and barred states within a scale length. Both models become more concentrated with time. As we shall see below, despite the similarities in the circular velocity curves, the dynamics of the two models are different. Therefore, circular velocity curves on their own do not control bar dynamics.

\begin{figure} \centering \includegraphics[width=3.in]{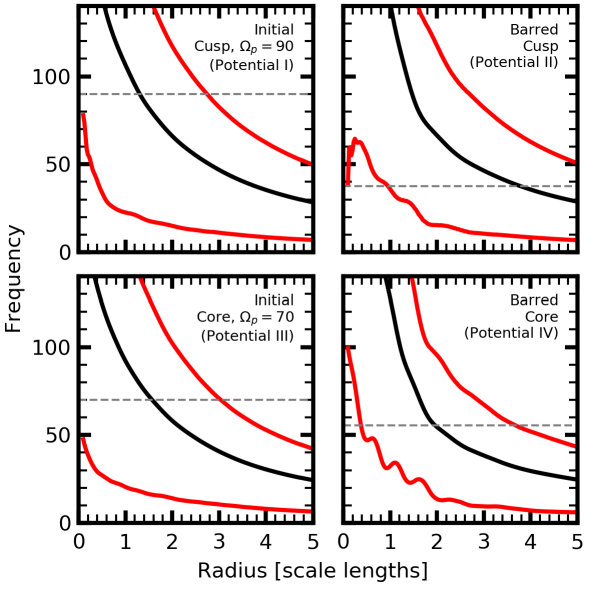} \caption{\label{fig:lindblad} Frequency versus radius (in disc scale lengths) for the four galaxy fixed potentials. In the barred potentials, we compute the frequency using the epicyclic approximation computed along the bar major axis. The black lines plot $\Omega$, which indicates corotation (CR); the lower red lines are $\Omega - \frac{1}{2}\kappa$, the inner Lindblad resonance (ILR) and the upper red lines are $\Omega + \frac{1}{2}\kappa$, the outer Lindblad resonance (OLR). The dashed grey line is the measured (assumed) pattern speed for the barred (initial) potentials.} \end{figure}

\subsubsection{Characterising the bar} \label{subsubsec:bardetermine}

A crucial ingredient in the algorithm that identifies trapped orbits is the phase angle of the bar. Previously we had employed ellipse fitting \citepalias{petersen16a}, the most traditional bar determination metric, or stellar surface density Fourier analysis. Both of these methods are subject to large-scale contributions that may not be related to the self-gravitating bar, such as spiral structure.  Ideally, one would be able to identify the sine and cosine terms of a Fourier analysis with some sort of spatial filter that eliminated the large-scale structure. Fortunately, the harmonic basis itself provides a spatial representation of the $m=2$ power. A harmonic method for determining the size and pattern speed of the bar is more robust than an ellipse fitting method, which is biased by the selection of bar metrics, such as the chosen ellipticity where the bar ends.

We reduce the contamination to the bar phase angle from spiral features in the outer disc by selecting the $m=2$ radial functions that best represent the bar. We have verified that the $m=n=2$ function produces the best characterisation of the bar pattern speed for all the models studied in this work. A comparison with the $m=2,~n=1$ harmonic shows that the position angle varies only modestly from the $m=n=2$ harmonic. However, the larger-scale $n=1$ term also contains some spiral-arm power, hence our choice to use $m=2,~n=2$.

\subsubsection{Figure rotation} \label{subsubsec:figure_rotation}

The dynamics and orbital structure are driven by the pattern speed of the bar, $\Omega_p$. The rotation of the model introduces the Coriolis and centrifugal forces in the bar frame, which depend on $\Omega_p$. For the barred potential models, we determine $\Omega_p$ by calculating finite differences in the rate of change of the coefficient phase in a finite window of the time series of coefficients from the self-consistent simulation. We calculate the uncertainty in $\Omega_p$ to be 5 per cent. Fortunately, we find that variations of 5 per cent to the pattern speed make little difference to the resultant orbital structure. For the initial potentials, Potentials I and III, we test two pattern speeds: $\Omega_p=0$, which reveals the unperturbed structure of the disc and halo system, and an estimated $\Omega_p$ from the self-consistent simulation. We estimate $\Omega_p$ using the coefficient phases as above for the earliest time when $\Omega_p$ can be measured: $T\approx0.2$. For the initial cusp we use $\Omega_p=90$ and for the initial core we use $\Omega_p=70$. We apply these pattern speeds to the $T=0$ potential models below. As we shall see, the introduction of figure rotation, and thus Coriolis and centrifugal forces, reveals that orbital structure varies with $\Omega_p$.

In Figure~\ref{fig:lindblad}, we show the primary resonance locations from epicyclic theory using the gravitational potential along the bar major axis for the measured bar pattern speed. The left panels plot the initial potentials (Potentials I and III) and we see that lowering the assumed pattern speed moves the calculated corotation radius outward. We also observe that the ILR does not exist at all in the initial core model for all realistic values of $\Omega_p$. The outer Lindblad resonance (OLR) exists for all values of $\Omega_p$. However, in the barred cusp, the radius of the OLR occurs at such large radii (and thus low stellar density) so that it would have little influence on the structure of the disc. In the right panels, we plot the barred potentials (Potentials II and IV) and compute the frequency by using $\Phi(r)$ along the major axis of the bar using the epicyclic approximation. As we shall see below, the potential along the major axis of the bar better estimates the frequencies in a self-consistent simulation than any other strategy. Calculating frequencies using $\Phi(r)$ along the minor axis of the bar results in a shallower potential profile, and thus the resonances move inward. In both the right panels, we assume the calculated pattern speed to estimate the location of the key resonances. The presence of the bar, despite the increased concentration of mass (obvious in the changed circular velocity curve at $r<2R_d$, cf. Figure~\ref{fig:circular_velocity}), results in the location of the key resonances occurring at larger radii than in their initial counterparts. The bar perturbation and the mass rearrangement resulting from secular evolution create an ILR in the barred core potential (Potential IV) where none existed in the initial core case (Potential III), as well as creating a second ILR at a larger radius in the barred cusp potential (Potential II).

\begin{figure*} \centering \includegraphics[width=7.0in]{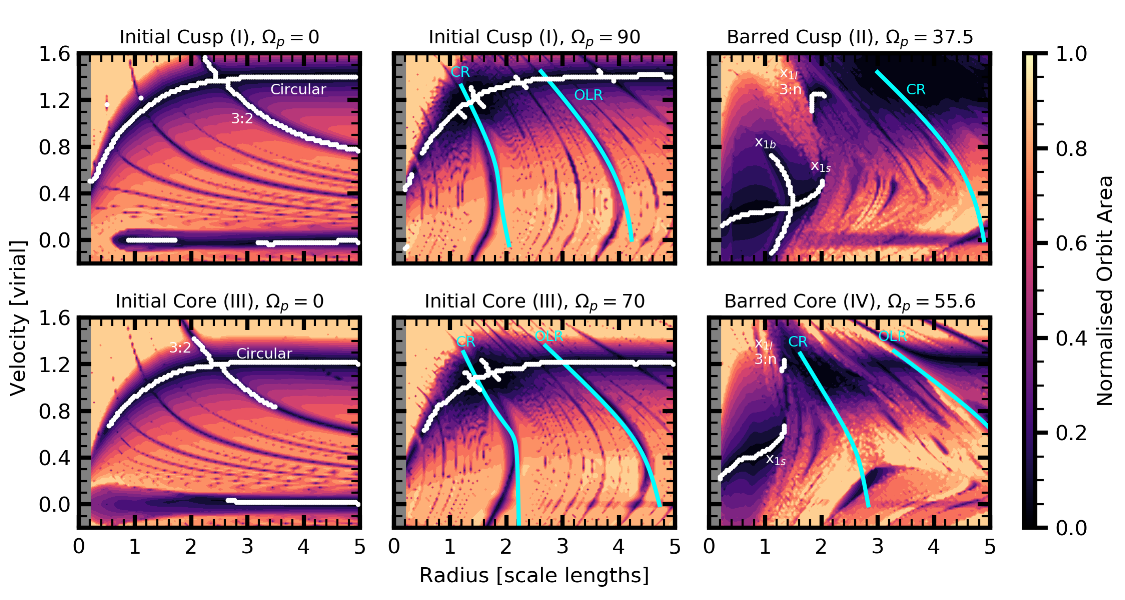} \caption{\label{fig:fishbone} Orbit-area diagrams as a function of radius and tangential velocity, where only orbits with $\dot{R}=0$ along the bar major axis are shown. We include two values of $\Omega_p$ for each of the initial models. The colour map runs from 0 (an orbit which is closed) to 1 (an orbit which completely fills the area of a circle with radius $\rturn$). In each panel, we highlight and label key commensurabilities identified with the tessellation algorithm in white. We plot and label the locations of corotation and the outer Lindblad resonance, as computed numerically from the monopole where possible, in cyan. The commensurabilities are discussed in detail in Section~\protect{\ref{subsec:cuspmodel}} for the cusp model and Section~\protect{\ref{subsec:coremodel}} for the core model. The grey region at $0.0<R_d<0.2$ was not integrated, owing to the limits of numerical resolution in this study. Some orbits may be shown twice; once at $\rapo$ and once at $\rperi$, reflecting where they may be observed along the bar major axis.} \end{figure*}

\subsection{Orbit Atlas Construction} \label{subsec:orbit_atlas}

We detail the construction of an atlas of orbits for the four fixed galaxy potentials described above. The atlas consists of a time-series of orbits for each model with a range of starting points. In section~\ref{subsubsec:initial_conditions}, we describe the starting points for these orbits. In section~\ref{subsubsec:integration} we describe both the procedures for orbit integration and the details of our implementation.

\subsubsection{Orbit starting point selection}\label{subsubsec:initial_conditions}

In \citetalias{petersen16a} we describe phase space using energy and angular momentum. Angular momentum was expressed as a fraction of the angular momentum of a circular orbit at the same radius, i.e. $X\equiv L_{z, {\rm orbit}}/L_{z, {\rm circular}}$ to create a roughly rectangular grid that extended from radial to circular obits. However, for an analysis of a strongly non-axisymmetric system, the use of axisymmetric orbit quantities (e.g. $L_{z, {\rm circular}}$) is inappropriate. For this work, we choose a more observationally-motivated set of dimensions: turning point radius ($\rturn$) and turning point tangential velocity ($\vturn$), where $\dot{R}=0$. One may still find a dominant closed-orbit family that is analogous to the circular orbit in an axisymmetric case, and we define $V_{\rm closed}$ at any given radius, which we use as a signpost for interpreting the atlas of orbits.

In a real system, orbits will typically be observed close to apocentre. Therefore, we will generally discuss $\rapo$ and $\vapo$ for orbits, and restrict our discussion to those orbits where $\vturn=\vapo$ and $\rturn=\rapo$ along the bar major axis. We construct the atlas by integrating orbits in a potential obtained from the BFE at a particular time. We apply the bar pattern speed, as described in section \ref{subsubsec:figure_rotation}, to $m>0$ subspaces. We release orbits with turning points along the major axis of the bar potential. We have investigated other release angles, but find that the bar axis is the most illustrative of the dynamics. In certain cases, it is necessary to use off-axis release angles to find orbits that are known to be relevant (described below), but we do not perform an exhaustive search of parameter space. We reserve a detailed study of the off-axis release angles for future work. While our orbit atlas does not fully sample phase space, this `pseudo-phase-space' gives a intuitive understanding of the system, and can be directly applied to observations.

For this study, we also restrict orbits to the disc midplane for simplicity. The inclusion of non-planar motion would be straightforward, although the phase-space is complex to explore. We will investigate vertical commensurabilities in future work. We choose to uniformly sample the $\rturn-\vturn$ plane. Our atlas makes no attempt to follow the phase-space distribution of our galactic models. So in principle, the same orbit could be generated both at $\rperi$ and at $\rapo$.  However, this ambiguity does not affect our goal of describing the structure spanned by the allowed orbit families. Rather, it allows us to compare the initial orbits with orbits in strongly non-axisymmetric systems.  Similarly, the atlas includes phase space that may not be occupied in the simulations (e.g. nearly radial orbits at large galactic radii). We choose $\rturn \in (0.2R_d, 5R_d)$ and $\vturn \in (-0.2\vvir, 1.6\vvir)$. Our limit of $\rturn>0.2R_d$ restricts our analysis to regular orbits. We will return to the question of central orbits in future work. Our limit of $\rturn<5R_d$ restricts our analysis to the region that contains all of the interesting commensurabilities.

We wish to explore the entirety of the relevant phase space and from an initial study of the simulations we see that retrograde orbits play some role in the dynamics of the disc at small radii. Hence, we truncate $\vturn$ at $\vturn=-0.2\vvir$ to study the relevant retrograde phase space. Similarly, at large $\vturn$, some orbits may occasionally be driven to large velocities at a given radius by a non-axisymmetric potential, joining new orbit families.

\subsubsection{Integration} \label{subsubsec:integration}

In the rest of this section, we describe specific details of our integration scheme, based on the leapfrog integrator used in {\sc exp}. Our integrator includes the following features beyond the standard leapfrog integrator: (1) an implementation of adaptive timesteps from {\sc exp}, described in \citetalias{petersen16a}, with minimum timestep thresholds; (2) completion criteria set by either total integrated time or the number of apoapses encountered. We use the minimum timestep in the self-consistent simulations, $dt_{\rm vir} = 3.2\times10^{-5}T_{\rm vir}$. We truncate the evolution after 50 radial periods have been completed or a maximum of $\Delta T=0.64 T_{\rm vir}$.

We define each component with a unique set of basis functions. The orbit integration may use the full potential from the simulation or a subset of basis functions for computational efficiency and accuracy. By excluding higher-order terms that do not influence the integration of individual orbits, we can achieve $\frac{n-n'}{n}$ or $1-\frac{l'^2}{l^2}$ per cent speedups, where $n$ ($l$) is the total number of radial (azimuthal) halo functions and $n'$ ($l'$) is the number of retained radial (azimuthal) halo functions. After inspecting the signal-to-noise ratio in the coefficients, we choose not retain higher order halo azimuthal terms with $l>2$, resulting in an 88 per cent speedup of the halo calculation, without any significant differences in the results.

We leave our integration flexible in the following ways: (1) the number of azimuthal harmonics in the disc may be specified at run time, which allows for restriction to the monopole contribution only and of eliminating odd harmonics; (2) the range of radial basis functions, which allows for testing the role of low signal-to-noise coefficients; and (3) varying the bar pattern speed. We do not apply odd-order azimuthal harmonics, which are empirically determined in the self-consistent simulations to have a different pattern speed than the even multiplicity azimuthal harmonics. In principle, we could use different values of $\Omega_p$ for individual harmonic orders, e.g. $\Omega_{p,~m=1}$ and $\Omega_{p,~m=2}$, allowing for an investigation of the dipole's influence separately from that of the quadrupole. We aim to study this phenomena in future work.

\subsection{Tessellation Algorithm}\label{subsec:tesselation_algorithm}

We use Delaunay triangulation (DT) to compute the physical volume that an individual orbit occupies, transforming a discrete time-series of $(x,y,z)$ to a volume. For our restriction to the disc plane in this study, the two-dimensional DT yields an area. We have tested three-dimensional DT, and will make vertical commensurabilities that are revealed in three dimensions the focus of future work.

We use the DT algorithm with the triangle heuristic described in Appendix~\ref{appendix:tesselationalgorithm} to compute the orbit areas, $A$, in the $\rturn-\vturn$ plane described in Section~\ref{subsubsec:initial_conditions}.  We refer to this as the {\it closed-orbit map}, for its utility in identifying closed orbits. The loci of $A\approx0$ defines orbit families, which we refer to as {\it valleys}. Valleys may be strong (wide valleys with large regions of $A\approx0$) or weak (narrow valleys with only a small path satisfying $A\approx0$). The valleys provide a skeleton of the orbits in a given potential, tracing the commensurate orbits that support the structure of the galaxy model. We, therefore, refer to the figures that show the orbit area at each point in the $\rturn-\vturn$ plane as {\it orbital skeletons}.

The identification of commensurate orbits provides an important theoretical link between a perturbation theory interpretation and fully self-consistent simulations \citep{contopoulos80, tremaine84, weinberg07a, weinberg07b} to find and describe trapped orbits.

\begin{figure*} \centering \includegraphics[width=6.5in]{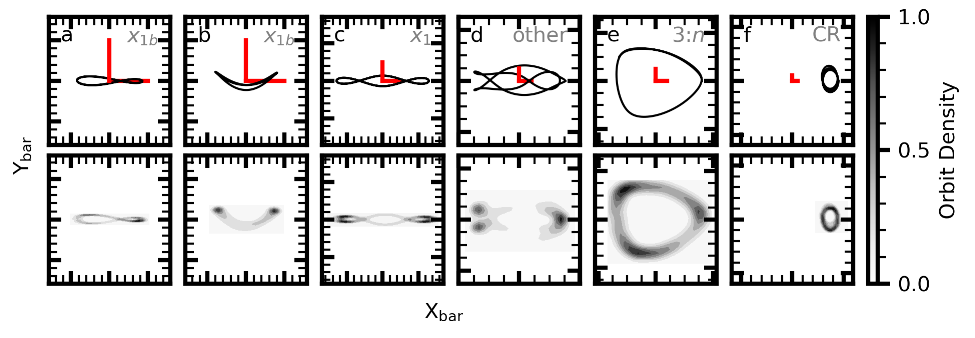}\caption{\label{fig:orbmap1} Six example integrations from the Barred Cusp (II) model. In each column, we show the trajectory (upper panel) and time-integrated density (lower panel). Panels a and b show orbits from the $x_{1b}$ family, where panel a is a `symmetric' $x_{1b}$ orbit (an `infinity' orbit), and panel b is `asymmetric' (a `smile' orbit). Panel c shows a strong $x_1$ orbit. As it is a shorter period $x_1$ than another $x_1$ subfamily at this value of $\rapo$, we call this an $x_{1s}$ orbit. Panel d shows a derivative of an $x_1$-like orbit with higher-order structure and a long period, which is an `other' orbit in our classification. Panel e shows a 3:2 orbit. Panel f shows a corotation (CR) orbit. In each of the upper panels, the red scale bar is half a disc scale length. All orbits are plotted in the frame rotating with the bar.} \end{figure*}

\section{Fixed potential study results}\label{sec:results}

We apply the tools described above to the potentials described in Section~\ref{subsec:fixedpotentials} with the goal of locating and identifying the orbit families in each. Closed-orbit maps for the six orbit atlases calculated from the four potential models are shown in Figure~\ref{fig:fishbone}.  Some orbits are shown twice; once at $\rapo$ and once at $\rperi$, if the orbit is released with $\vturn>v_{\rm closed}(\rturn)$. We leave these orbits on the diagram to facilitate later comparison with self-consistent models. The colour map shows the values of the area $A$, as sampled by the initial conditions listed above. The colour scheme is uniform throughout the paper; colours in Figure~\ref{fig:fishbone} may be compared to the two orbits in Figure~\ref{fig:example_orbits} for intuition on the colour map. The white lines in Figure~\ref{fig:fishbone} are the identified valleys. We do not plot all the valleys identified, but rather restrict ourselves to those with dynamical relevance to bar-driven evolution to avoid confusion. Further, for radii beyond the strong influence of the non-axisymmetric bar force, we use the monopole-calculated frequency to estimate the location of CR and OLR. Near the bar, the non-axisymmetric distortions are too large for resonances to be calculated from the monopole.

The valleys track the commensurabilities throughout the bar region, demarcating orbit families.  The particular family may be precisely identified by inspecting orbits from each valley. In many cases, the valleys intersect. At these points, we expect to find weakly chaotic behaviour in a self-consistent simulation. We focus on regions in phase space around the commensurabilities to study the details of secular evolution, such as angular momentum. Overall, we find similar orbit families to classic analytic studies (e.g. \citealt{contopoulos89, athanassoula92, sellwood93}). Figure~\ref{fig:fishbone} bears some resemblance to the so-called `characteristic diagram' found in the analytic literature, although an additional advantage of the closed-orbit map is that the steepness of the valley indicates the measure of orbits that resemble the parent; broad low-$A$ valleys imply the existence of many nearly commensurate orbits. The regions of low area in the closed-orbit map (i.e. $A<0.02$) describe the range of librating orbits in $\rturn-\vturn$ space that resemble the parent orbit (Figure~\ref{fig:fishbone}).

We first discuss the orbit families in the cusp models before turning to the cored models. For both sets of models, we begin with the non-rotating axisymmetric models to describe the unperturbed commensurabilities (Sections~\ref{subsubsec:ecusp0} and \ref{subsubsec:ecore0}). We then impose a pattern speed upon the axisymmetric models (Sections~\ref{subsubsec:ecusp90} and \ref{subsubsec:ecore70}), followed by the bar-like non-axisymmetric models in Sections~\ref{subsubsec:bcusp} and \ref{subsubsec:bcore}. In Section~\ref{subsec:selfconsistent}, we present the results of applying the tessellation algorithm to orbits extracted from the self-consistent simulation. We compare the differences between the fixed potential models and the self-consistent simulations in Section~\ref{subsec:results_summary}.

\subsection{Cusp Models} \label{subsec:cuspmodel}

\subsubsection{Initial Cusp (I), $\Omega_p=0$}\label{subsubsec:ecusp0}

The zero pattern speed initial cusp potential (Potential I) reveals the commensurate families for a disc embedded in a dark matter halo. We present the closed-orbit map in the upper left panel of Figure~\ref{fig:fishbone}. We overlay the orbital skeleton as determined via our tessellation algorithm.

The circular orbit curve (labelled) is the most clearly defined valley. Orbits above this curve will be at pericentre, while orbits below this curve will be at apocentre. Crossing the circular orbit valley are several $m$:$n$ commensurabilities, where $n$ is the radial order and $m$ is the azimuthal order, satisfying equation~(\ref{eq:resonances}). The 3:2 commensurability (labelled) is the strongest commensurability crossing the circular orbit curve. Multiple 3:$n$ families exist in barred systems, These include the 3:1 family, which has been previously studied in the literature \protect{\citep{athanassoula92}}, and is considered a bifurcation of the $x_1$ family. Some of the 3:$n$ overlap with the $x_1$ families; this overlap is a channel for bar growth \citep[as described in][hereafter Paper II]{petersen18b}. The values of $m$ increase toward smaller radius, such that the next strongest commensurability curve is 5:2, then 7:2, and so on (not labelled). These high-$m$ and $n$ resonances are not expected to be important for the evolution of the system. We will confirm this in Section~\ref{sec:selfconsistent}. A physically uninteresting radial orbit commensurability valley also appears at $\vapo=0$.

\subsubsection{Initial Cusp (Potential I), $\Omega_p=90$}\label{subsubsec:ecusp90}

The rotating initial cusp, also with the underlying Potential I, has structure not present in the non-rotating version of the potential for a pattern speed of $\Omega_p=90$, an estimate for the initial formation pattern speed of the bar (middle left panel of Figure~\ref{fig:fishbone}). Owing to the axisymmetric nature of the potential, the circular orbit valley is unchanged from the same potential model with $\Omega_p=0$. However, the radial orbit commensurability seen at $\vapo=0$ in the non-rotating case is not well defined in the rotating model, occupying a negligible region of phase space that is below the resolution of the closed-orbit map.

In preparation for using this technique to investigate the dynamics in a rotating non-axisymmetric potential, we now describe the commensurate structure in the rotating frame. A rotating model admits the familiar low-order resonances, including the inner Lindblad resonance (ILR), corotation (CR), and the outer Lindblad resonance (OLR). Many higher-order resonances are clearly seen as low-area (dark) loci. These features correspond to the higher order resonances discussed above for the nonrotating model. They are unlikely to be important in a time-varying potential where the pattern speed and underlying potential changes faster than the orbital time for a high-order closed orbit.

\subsubsection{Barred Cusp (Potential II)}\label{subsubsec:bcusp}

The non-axisymmetric barred cusp model (Potential II) has orbit families not present in the axisymmetric models.  We classify three subfamilies of $x_1$ orbits: \begin{enumerate} \item $x_{1b}$ orbits, which may be symmetric or asymmetric about the axis perpendicular to the bar. We show a symmetric `infinity'-sign-shaped orbit in panel a of Figure~\ref{fig:orbmap1}, and we show an asymmetric `smile'-shaped $x_{1b}$ orbit in panel b (the asymmetric orbits are an example of an orbit which is more readily identified from an off-axis release)\footnote{`Symmetric' in the case of $x_{1b}$ orbits refers to symmetry across the axis perpendicular to the bar in papers on $x_{1b}$ orbits \protect{\citep[e.g.]{contopoulos89}} . Thus panel a of Figure~\protect{\ref{fig:orbmap1}} is a `symmetric' $x_{1b}$ orbit, owing to the $y$-axis symmetry, and panel b is asymmetric. Without any figure rotation, the symmetric $x_{1b}$ orbit looks exactly like an infinity sign, that is, the crossing point is centred rather than off-centre, the `antibanana' of \protect{\citet{miralda89}}.}. \item $x_{1s}$ orbits, short-period bifurcated standard $x_{1}$ orbits (with `ears', panel `c' of Figure~\ref{fig:orbmap1}). \item $x_{1l}$ orbits, long-period elongated $x_1$ orbits. \end{enumerate} We morphologically classified the families identified in Figure~\ref{fig:fishbone} through a visual inspection of the orbit atlas. While many $m$:$n$ orbits with $m>1$ are clearly observed in Figure~\ref{fig:fishbone}, we choose to mark only $m=3$. Several low-order even $n$ orders comprise the $m=3$ feature, and are indistinguishable in $\rturn-\vturn$ at the resolution of Figure~\ref{fig:fishbone}. This is labelled as 3:$n$ in Figure~\ref{fig:fishbone}. In panel `e' of Figure~\ref{fig:orbmap1}, we plot an example 3:2 orbit. All orbits that are asymmetric across the $x$-axis in Figure~\ref{fig:orbmap1} have corresponding mirror image orbits, where an orbit with one symmetry leads the bar pattern and an orbit with the other symmetry trails it. With a fine enough grid, we find arbitrarily high order commensurabilities (see, e.g. the unidentified structure in Figure~\protect{\ref{fig:fishbone}} from the 3:$n$ position to CR and beyond). In this work, we restrict our analysis to the low-order strong commensurabilities that form the persistent orbital structure of the barred galaxy.

In panel f of Figure~\ref{fig:orbmap1}, we show an example CR orbit in the barred cusp potential (Potential II), which has a strong CR feature. CR is the lowest-order resonance present in the model, with wide-ranging dynamical effects for secular evolution discussed extensively in the literature (see \citealt{sellwood14} for a review). CR orbits are particularly easy to recover using the tessellation algorithm owing to the tiny area spanned by their trajectory, evident in pane f of Figure~\ref{fig:orbmap1}. Owing to the slow precession rates for orbit apoapses near CR in this model, the orbital-skeleton-tracing algorithm described in Appendix~\ref{appendix:tesselationalgorithm} identifies large regions with $A<0.1$ and shallow slopes of $\partial A/\partial V$ and $\partial A/\partial R$, making tracing valleys ambiguous. This is not a limitation of the algorithm, per se, but rather results from the low $A$ values in the weakly perturbed, nearly circular outer disc. We, therefore, opt to include only the monopole-derived commensurabilities at radii outside of the bar radius.

\begin{figure} \centering \includegraphics[width=2.9in]{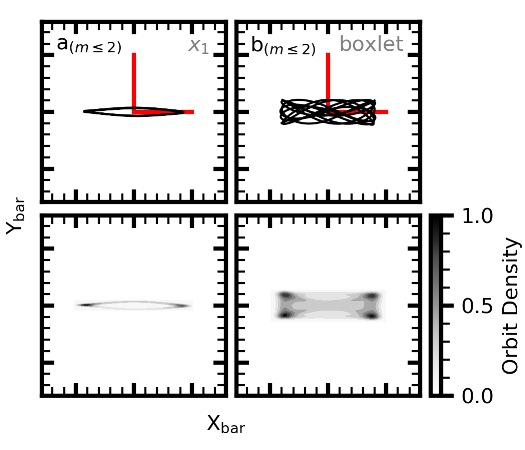} \caption[Orbits integrated in a limited $m=2$ bar potential.]{\label{fig:orbmap1_M4} Left column: the same orbit as in panel `a' of Figure~\protect{\ref{fig:orbmap1}}, integrated applying only up to the quadrupole potential disturbance ($m\le2$). The orbit is now a standard $x_1$ orbit. Right column: the same orbit as in panel `b' of Figure~\protect{\ref{fig:orbmap1}}, also integrated applying only up to the quadrupole potential. The orbit is now a rotating boxlet. In both upper panels, the red scale bar is half a disk scale length. All orbits are plotted in the bar frame.} \end{figure}

Higher-order azimuthal harmonics of a strongly non-axisymmetric gravitational potential can affect the shape and stability of orbits in low-order resonance. New families appear owing to relatively small but important changes in the potential. For example, the exclusion of the $m>2$ harmonics from the barred cusp potential model (II) results in the disappearance of the $x_{1b}$ family. Inspection of all orbits that are part of the $x_{1b}$ family when $m\le 6$ reveals that the $x_{1b}$ track no longer exists when we restrict the potential to $m\le2$. However, this should not be interpreted as evidence that $2<m\le 6$ causes new resonant structure into which the $x_{1b}$ orbits are trapped, but rather that $2<m\le 6$ distorts the potential shape allowed by the quadrupole only into a potential that admits $x_{1b}$ orbits.  In Section~\ref{sec:selfconsistent}, we will see that $x_{1b}$ orbits are important for growing the bar in length and mass. In Figure~\ref{fig:orbmap1_M4}, we integrate the same orbits as in panels a and b of Figure~\ref{fig:orbmap1}, except that we limit the harmonics included in the potential to $m\le2$. The orbits are no longer $x_{1b}$ orbits. The infinity morphology $x_{1b}$ orbit is now a part of the $x_1$ family. The smile morphology $x_{1b}$ orbit has become a `rotating boxlet' orbit\footnote{Some orbits do not show any apparent structure in the inertial frame, filling in an entire circle, but appear to be rectangular `boxes' in the rotating frame owing to the inner quadrupole of the bar ($\Phi_{\rm bar}\propto r^2$) approximating a harmonic potential. The non-rotating version of these orbits have been called boxlets \protect{\citep{miralda89, lees92, schwarzschild93}}. We therefore call these `rotating boxlets'. The maximum radial extent of rotating boxlets informs the structure of the potential, but is reserved for future work studying the innermost regions of the potential models.}.

This finding also suggests that $m=2$ parametrisations of the MW bar\footnote{Ellipsoid-derived bar models such as the Ferrers bar \protect{\citep{binney08}} will naturally admit $m=4$ power, depending on the axis ratio, such that an increase in axis ratio will increase the $m=4$ power relative to $m=2$. Additionally, the density profile of the bar will contribute to the $(m=4)/(m=2)$ ratio, with an increase in central density leading to a lower $(m=4)/(m=2)$ ratio.}, such as those derived from the potential of \citet{dehnen00}, \citep[e.g.][]{antoja14, monari16, monari17a,hunt18b}, may entirely miss important families of orbits, {\it even if the orbits do not appear to exhibit four-fold symmetry}. Other recent models for the MW have suggested the importance of the $m=4$ component of the bar for reproducing the observed velocities near the Sun \citep{hunt18a}.

In the barred cusp potential (Potential II), CR intersects the closed orbit track at $\rturn=3.1R_d$. Owing to the relatively slow pattern speed, $\Omega_p=37.5$, CR is at a fairly large radius in this potential model; if we assume that the radial terminus of the $x_{1s}$ family is the length of the bar, then the ratio of corotation to the bar length in this model is $\mathcal{R}=\frac{3.1R_d}{2.1R_d} = 1.47$, well within the `slow' regime for bars \citep{athanassoula92}. The 50 per cent pattern speed slowdown since formation has implications for the observed fast bar-slow bar tension \citepalias[see e.g. ][]{petersen16a}, suggesting that a direct observation of the $x_{1s}$ radial terminus would prove a powerful diagnostic for determining the pattern speed of bars (see Section~\ref{sec:observations}). Choosing a smaller (larger) pattern speed than the measured one causes the location of the vertical extension of $x_{1s}$ orbits in $\vapo$ to move outward (inward) in radius. Artificially decreasing (increasing) the amplitude of the $m>0$ components of the model potential causes the horizontal $\rturn$ locus of the $x_{1s}$ orbit valley to drift to lower (higher) velocities, before the $x_{1s}$ valley ultimately settles at $\vapo=0$ in the case of an axisymmetric disc.  The $x_1$ track is related to radial orbits in a non-rotating triaxial model \citep{valluri16}.

Lastly, a comment about the ultra-harmonic resonance (UHR), which occurs when $\Omega_p = \Omega_\phi - \Omega_r/4$. While simulations reportedly are able to detect the UHR \citep{ceverino07}, we do not find any evidence for UHR orbits in our simulations in the form of a valley where one might roughly expect to see the UHR. Indeed, in the fixed potential analysis of the barred cusp model, a gap in the $x_1$ valley appears with significantly nonzero orbit area. Analytic work \citep{contopoulos88} suggests that the UHR can arise as a continuation of the $x_{1s}$ orbits we do observe. Inspection of the orbit trajectories, despite their nonclosure, supports this conclusion: we find orbits that appear to be near UHR. Thus, we appear to see evidence for UHR effects in the fixed barred cusp potential, although not in the form of a detectable closed orbit.

\subsection{Core Model} \label{subsec:coremodel}

\subsubsection{Initial Core (Potential III), $\Omega_p=0$}\label{subsubsec:ecore0}

The closed-orbit map for the nonrotating initial cored model shares many similarities with the nonrotating cusp model (Potential I). A comparison of the cusp (I) and cored (III) nonrotating models -- the left panels of Figure~\ref{fig:fishbone} -- shows the effect of the halo model on the zeroth order commensurabilities from a disc embedded in the halo. Despite the nearly identical nature of the disc monopole, as shown by the similarity in the disc component of the circular velocity curve in Figure~\ref{fig:circular_velocity}, the contrast in halo model monopole acts to create a significantly different total circular velocity within a few scale lengths. In particular, the peak of the circular velocity curve drops from $\vapo=1.4\vvir$ in the cusp model to $\vapo=1.2\vvir$ in the cored model. The location of the $m$:$n$ commensurabilities shift to smaller radii for all values of $\vapo$. We mark the strong 3:2 commensurability and note that the other obvious tracks are equivalent to the tracks in the nonrotating cusp model.

\subsubsection{Initial Core (Potential III), $\Omega_p=70$}\label{subsubsec:ecore70}

We plot the closed-orbit map for the rotating initial core model in the lower left panel of Figure~\ref{fig:fishbone}. While we applied a lower imposed pattern speed than for the initial cusp model, estimated from the self-consistent core simulation, we see similar structure in the rotating initial core model (III) to that of the rotating initial cusp (I). Corotation reaches larger values of $\rturn$ than those reached in the rotating initial cusp model, reaching a maximum at $\rturn=2.1R_d,~\vturn=0.4\vvir$. We observe a strong 3:2 commensurability at larger values of $\rturn$ compared to the rotating initial cusp model (I). CR and OLR computed from the monopole intersect the circular velocity curve at positions more or less in agreement with the estimate from Figure~\ref{fig:lindblad}.

\subsubsection{Barred Core (Potential IV)}\label{subsubsec:bcore}

While many of the major commensurabilities remain, albeit at different locations in $\rturn-\vturn$ space, we do not observe the $x_{1b}$ family. As in the barred cusp model, we see both $x_{1s}$ and $x_{1l}$ orbits, where the $x_{1l}$ orbits again are in the same location in $\rturn-\vturn$ as the $m=3$ series of commensurabilities\footnote{While not a formal phase-space, as discussed elsewhere in this work, residing near the same location in the orbit atlas $(\rturn,\vturn)$ implies that orbits and families must be adjacent in phase-space as well, as the mapping to $(E,\kappa)$ is unique.}. Despite the models having approximately identical disc monopoles (cf. Figure~\ref{fig:circular_velocity}), the difference in the halo model and the larger pattern speed of the bar, $\Omega_p=55.6$, results in CR being located at a significantly smaller radius than in the barred cusp model (II). CR intersects the dominant closed-orbit curve (analogous to the circular orbit curve in the axisymmetric potential) at $\rturn=1.8R_d$, compared to $\rturn=3.1R_d$ for the barred cusp.  Similarly, OLR appears at a smaller radius when compared to the barred cusp model (II). Comparing CR with the maximum radial extent of the $x_{1s}$ orbits at $\rturn=1.4R_d$, the ratio of CR to bar length ends up as $\mathcal{R}=1.28$, within the fast bar regime, versus $\mathcal{R}=1.47$ for the barred cusp (II).

\subsection{Fixed-Potential vis-a-vis Self-Consistent Simulations} \label{subsec:selfconsistent}

\begin{figure} \centering \includegraphics[width=3.2in]{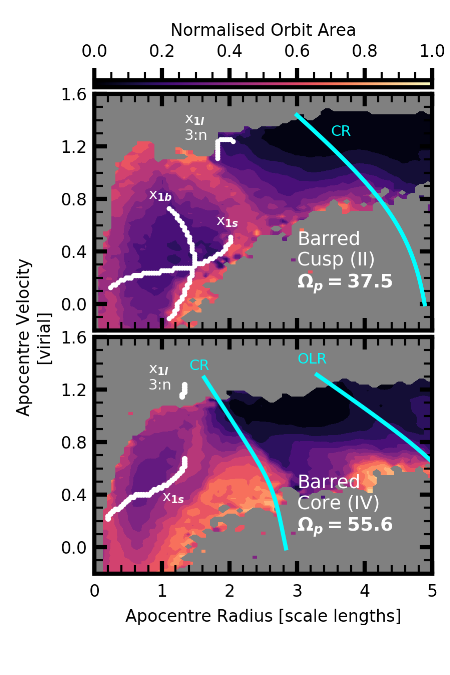} \caption{\label{fig:live_quintile} Orbit area as a function of $\rapo$ and $\vapo$ for live simulation orbits from $1.8<T<2.2$, computed from the lowest decile for each bin in $\rapo$ and $\vapo$. Regions without sufficient orbit sampling (i.e. fewer than ten orbits in a bin) are greyed out. The upper panel shows orbits from the barred cusp model (II) and the lower panel shows orbits from the barred core model (IV). The commensurability tracks for the corresponding model from Figure~\protect{\ref{fig:fishbone}} are over plotted in white, and monopole-determined commensurabilities are plotted in cyan.} \end{figure}

As discussed in Section~\ref{subsec:fixedpotentials}, both of the barred potentials are drawn from larger self-consistent simulations. In this section, we place the single snapshots back into the self-consistent context, using phase-space snapshots drawn from the self-consistent simulations to construct orbit-area maps. To retain similarity with the chosen fixed potential models (extracted at $T=2.0$), we use phase space outputs satisfying $T\in[1.8,2.2]$, with $dT = 0.002$, for a total of 200 outputs.

In Figure~\ref{fig:live_quintile}, we use the simulation phase space to generate self-consistent orbit area maps. We do this by first transforming the 200 outputs to a frame co-rotating with the bar, then feed the sequences for each orbit to the tessellation algorithm. An accurate orbit area requires a series of at least 2 orbital periods. Orbits with $\rapo\le5R_d$ generally satisfy this criterion. For each orbit, we then calculate its maximum $\rapo$ in the $\Delta T=0.4$ window and the corresponding tangential velocity at the maximum radius, $\vapo$. The orbits are put into rectangular bins in the $\rturn-\vturn$ diagram with $dr=0.1R_d$ and $dV=0.05\vvir$, based on their $\rapo$ and $\vapo$ values. For each bin, we calculate the lowest decile (10$^{\rm th}$ percentile) relative area from the distribution of relative areas found by the tessellation algorithm. We tested alternate particle selection criteria per bin, including mean, median, and minimum, and find that the lowest decile value provided an appropriate balance between feature extraction and overemphasis of outliers that appear in given bins owing to errors in determining $\rapo$ and $\vapo$.

The upper panel of Figure~\ref{fig:live_quintile} shows the results for self-consistent orbits drawn from the barred cusp model. Many regions in the $\rapo-\vapo$ plane are not populated in the self-consistent simulation (grey regions). Regions of low relative orbit area correspond to the commensurability tracks from Figure~\ref{fig:fishbone} superposed in white. The region at $\rapo<2R_d$ is dominated by nearly commensurate bar orbits close to the $x_1$ commensurability valley. However, the distribution is not symmetric in $\vapo$ around the $x_{1s}$ valley, but is biased to larger $\vapo$. A close inspection of the self-consistent simulation supports the existence of this bias, which exhibits orbits that resemble $x_{1b}$ orbits primarily leading the bar, expected for larger $\vapo$ orbits. A larger $\vapo$ at fixed $\rapo$ relative to the $x_1$ commensurability valley implies that the orbit has a larger angular momentum than the equivalent $x_1$ orbit, and the system is still evolving. At $\rapo=2R_d$ and $\vapo=1.2\vvir$, a valley appears that is not prominently seen in the tracks from Figure~\ref{fig:fishbone}. An artificially drawn extension of the 3:$n$ series commensurabilities following an isoenergy line would approximately account for this valley, which suggests that the fixed potential integrations may be missing some key ingredient that affects the self-consistent evolution. We discuss some possible explanations in Section~\ref{subsec:limitations}.

The lower panel of Figure~\ref{fig:live_quintile} shows the results for orbits drawn from the self-consistent barred core model. Once again, orbits with low relative orbit area gather near the $x_{1s}$ track. A second low relative orbit area feature, at similar $\vapo$ to the $x_{1s}$ track but at $\rapo\sim3R_d$, is also apparent. Comparison with fixed potential orbits reveals that this feature is probably an extension of the 3:$n$ orbits to lower energies. The relative prominence of features at below the closed orbit track in the self-consistent orbits when compared to the fixed potential orbits suggests that we have omitted some temporal coupling in the fixed potential integration, as we also found in the barred cusp model comparison above. Additionally, we find boxlets in the barred core model from inspection of the self-consistent simulation that are below the minimum radius of the orbit atlas ($\rturn<0.2R_d$) owing to the nearly harmonic potential resulting from the inner halo density profile.

\citet{aumer15} analysed an $N$-body simulation of a disc galaxy to compare it with the observed high $v_{\rm los}$ features in the MW observed from APOGEE, finding that orbits associated with the bar are able to reproduce the observed high $v_{\rm los}$ feature. Of these orbits, roughly half are on $x_1$ orbits. The others are on higher-order orbits . They attribute the orbits to dynamically cool, young stars formed outside the bar and captured by the growing bar. In their model, the feature is not visible at $|b|> 2$ deg. \citet{aumer15} was an improvement on the explanation of \citet{molloy15}, who was able to reproduce the signal when using $x_1$ orbits only, but not the entire galaxy.

\subsection{Summary} \label{subsec:results_summary}

The unique features of each orbit family tie the secular dynamical mechanisms to the underlying potential. Comparison between the fixed potential integration and the self-consistent simulations yields the following results: \begin{enumerate} \item We empirically locate CR and OLR, finding that the barred cusp potential (II) CR location is at a substantially larger radius than the barred core potential (IV) owing to the pattern speed rate of change and the size of the bar. \item We find the presence of the $x_{1b}$ orbit family in the barred cusp potential, but not in the barred cored potential (or in either of the axisymmetric models). \item The $x_{1l}$ and 3:$n$ commensurabilities are co-located at the end of the bar in the $\rturn-\vturn$ plane. \item Orbit families observed in the self-consistent simulation cannot be recovered without including the $m=4$ harmonic in the potential. \end{enumerate}

The most obvious difference between the barred cusp (II) and core (IV) potentials is the presence of the $x_{1b}$ track, which affects the structure of the orbits in the barred cusp model. Additionally, the 3:$n$ valleys lie in populated regions of phase-space in the barred cusp model. Inspection of the orbits in the self-consistent model shows that a common channel to add additional orbits the bar is the transition from 3:$n$, where $n=1$ or $n=2$, to $x_{1s}$ orbits over short timescales. This fuelling of the bar is the result of resonance passage that imparts a change in orbital actions, as per standard canonical perturbation theory. The close proximity of the $x_1$ and 3:$n$ families in $\rturn-\vturn$ raise the possibility that there might be actual resonance overlaps (formally heteroclinic connections) that may lead to orbit family switching. In either case, we observe that the 3:$n$ families serve as a conduit by which orbits can join the $x_{1l}$ family, which in turn can trade orbits with the $x_{1s}$ family\footnote{This mechanism is distinct from `strong chaos', as in \citet{chirikov79}. An example of a similar mechanism at work is `radial migration', a rapid process resulting from transient perturbations. See \citet{sellwood02}.}. No such channel exists in the barred core case as the phase-space region where the 3:$n$ and $x_{1l}$ orbit valleys intersect is not populated (cf. Figure~\ref{fig:live_quintile}). We will see in Section~\ref{sec:selfconsistent} that the lack of such a channel affects the evolution of a barred galaxy.

The fixed potential integrations presented here form a bridge between analytic potential study and fully self-consistent simulations. Canonical works on the dynamics of a disc and halo system have relied on the use of potentials with separability, such that the actions can be directly calculated (see, for example, \citealt{binney08}). Dynamical models that one fits to galaxies assume axisymmetry, which will disagree with the findings presented here. Both classic and modern MW potentials are largely assumed to be axisymmetric, in stark contrast to a multitude of observations indicating that the MW is strongly barred \cite[e.g.][]{blandhawthorn16}. The fixed potential analysis described here is a more precise tool for determining orbital structure and provides a complete picture of the orbital structure in the system.

\section{Application to Observations} \label{sec:observations}

\begin{figure} \centering \includegraphics[width=3.5in]{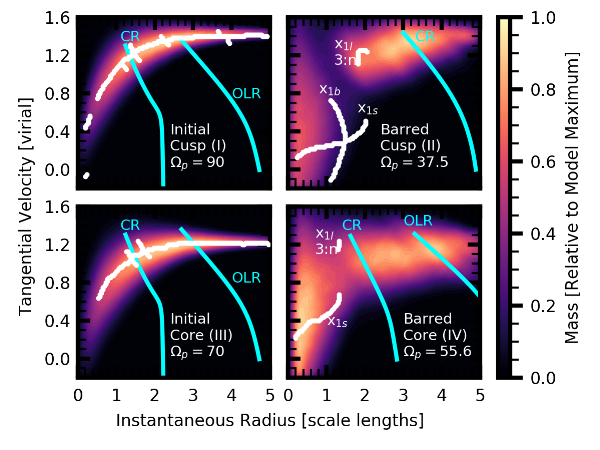} \caption{\label{fig:resonance_positions} Instantaneous tangential velocity, calculated as $v_T = (x\dot{y} - y\dot{x})/R$ versus instantaneous radius, drawn from the corresponding self-consistent simulation. The colour map is the mass relative to the maximum density at the same radius $R$ in the $(v_T,R)$ plane, for the four potential models. Overlaid on each are the traced commensurabilities from Figure~\protect{\ref{fig:fishbone}}, including assumed values of $\Omega_p$ for the initial models.} \end{figure}

We describe the tools presented in Sections~\ref{subsec:orbit_atlas} and \ref{subsec:tesselation_algorithm} in the context of observations and observational interpretation. The fixed potential analysis is useful for making direct inferences about the presence of different commensurabilities in observed galaxies. In a future application, we will train orbit finding algorithms to detect complex, sparsely sampled members of the orbit library.

We provide an instantaneous description of the phase space in $\rturn-\vturn$ coordinates to directly parallel observations made with integral field units (IFUs). In Figure~\ref{fig:resonance_positions}, we show the stellar mass as a function of instantaneous tangential velocity and instantaneous radius computed from the phase space distribution of the particles in the self-consistent simulation. To normalise the density map, we find the maximum density in binned $\rturn-\vturn$ space, set the value to be equal to 1 and scale all the other masses accordingly. In contrast to the discussion in Section~\ref{subsec:selfconsistent}, here we use only the instantaneous information from the phase space distribution (to facilitate comparison with images of galaxies). As orbits spend a larger fraction of their time near apocenter, the signal is not as diluted as one might fear, and hence we undertake a direct comparison of instantaneous quantities and apocenter quantities.

The upper panels of Figure~\ref{fig:resonance_positions} show the cusp potentials, with the initial cusp potential (I) on the left and the barred cusp potential (II) on the right. Overlaid on each of the panels in white are the corresponding commensurability skeletons from Figure~\ref{fig:fishbone}. The rotating initial cusp model (I) places corotation at a smaller radius, $r_{\rm CR}=1.3R_d$, than the majority of the disc mass\footnote{The maximum of the mass distribution in circular annuli is by definition at $r=2.2R_d$ for an exponential disc.}. Both the rotating and non-rotating initial cusp potentials admit a 3:2 frequency commensurability track that intersects the closed-orbit track at $r=2.6R_d$, as the 3:2 frequency valley results from the combined disc-halo mass distribution, independent of the bar.

The barred cusp potential (II, the upper right panel of Figure~\ref{fig:resonance_positions}) shows a number of features in the mass distribution that correlate with the commensurability traces. In particular, the mass associated with the bar resides within the maximum radius of the $x_{1b}$ track, with the material at the end of the bar ($r=1.8R_d$) at similar radii to the $x_{1l}$ and 3:$n$ orbits. The location of corotation is exterior to the majority of the disc mass. Local minima in mass density along the closed orbit track as one moves outward in radius appear to correlate with higher-order commensurabilities, suggesting that the orbit families in these regions are unstable.

For the initial core (III, the lower left panel of Figure~\ref{fig:resonance_positions}), like the initial cusp, the bulk of the mass distribution lies along the empirically-determined closed orbit track (shown in white). The spread in the measured tangential velocity values (at a fixed radius), reflects non-circular motions in the self-consistent simulation. As with the initial cusp, corotation is interior to the majority of the disc mass, at roughly the same radius, $r=1.5R_d$.

The barred core model (IV) is shown in the lower right panel of Figure~\ref{fig:resonance_positions}. Here, the contrast with the barred cusp model is striking. The mass distribution is more continuous between the eccentric bar orbits ($r<1.4R_d$) and the nearly circular orbits ($r>1.4R_d$). Without the presence of an $x_{1b}$ commensurability track, the bar is limited by the $x_{1s}$ track. Additionally, the $x_{1l}$ and 3:$n$ commensurabilities, despite residing at similar radii, do not appear to control the structure of the mass density. Rather, that role is ceded to corotation, which meets the closed orbit track at $r=2.0$, where a pile-up of orbits occurs. We again see that higher-order resonances (in this case part of the 5:$n$ series) cause a disruption in the closed orbit track at $r=3R_d$. Taken together, the barred cusp (II) and barred core (IV) models provide examples of commensurabilities that are responsible for structure in barred galaxies. In both models, the $x_1$ family dominates the structure of the bar itself, with the mass distribution apparently correlating with commensurabilities. The differences between the cusp and core models are significant enough to discern with an IFU and $\delta v\approx10~\kms$ velocity resolution targeting galaxies close enough to achieve $\delta r\approx 0.5$ kpc resolution.

The $\rturn-\vturn$ diagram may be constructed using IFU data for a range of inclinations ($20^\circ<i<70^\circ$) using a simple process: (i) deproject the galaxy and transform the line-of-sight velocity distribution to $x$ and $y$ velocities, with the $x$ axis aligned with the bar major axis, (ii) bin the position and velocity data by radius and velocity, (iii) plot the binned data in the $\rturn-\vturn$ plane. As shown in Figure~\ref{fig:resonance_positions}, the features associated with commensurabilities, e.g. the end of the bar in $\rturn-\vturn$ space, are discernible as a local minima in the $\rturn-\vturn$ distribution even without knowing the true apocenter values. The orbits spend proportionally more time at apocenter than at any other point in the orbit, leading to an effective weighted average, and reflect the closed orbits at resonances.

The $\rturn-\vturn$ diagram suggests that the role of CR is appreciably different between the barred cusp and core models. CR plays a larger role in driving evolution in the barred core model, where the resonance is in a well-populated region of phase space, but has little effect in the barred cusp model, where CR is located at a larger radius than the bulk of the disc material. Indeed, the large radius of CR in the barred cusp model suggests that it plays a minimal role in the evolution after the assembly of the bar at $T>1$ ($\approx 2$ Gyr scaled to the MW). The clearest diagnostic is the near-discontinuity in the mass distribution at the end of bar in the barred cusp model, while the barred core model maintains a track connecting bar and non-bar orbits.

Many of the orbits in the self-consistent simulation reside at considerable phase-space distance from commensurabilities; even in an apparently slowly evolving or steady-state evolutionary phase, orbits can be distant from their true parent orbit. Models that rely on the construction of mass distributions from a suite of orbits drawn from a particular simulation without time dependence or transients, such as Schwarzschild orbit superposition \citep{schwarzschild79, vandermarel98} and made-to-measure \citep{syer96, dehnen09, morganti12,portail15a} techniques could be biased against important orbital families that are short-lived, yet crucial for the dynamics, such as the $x_{1b}$ family. Therefore, we caution that this limitation (and others discussed in Section~\ref{subsec:limitations}) must be considered when interpreting the evolution from models that are constructed using closed orbit libraries. This may be particularly relevant for the MW, which shows evidence for continued dynamical evolution in the disc \citep{antoja18}.

\begin{figure*} \centering \includegraphics[width=6.5in]{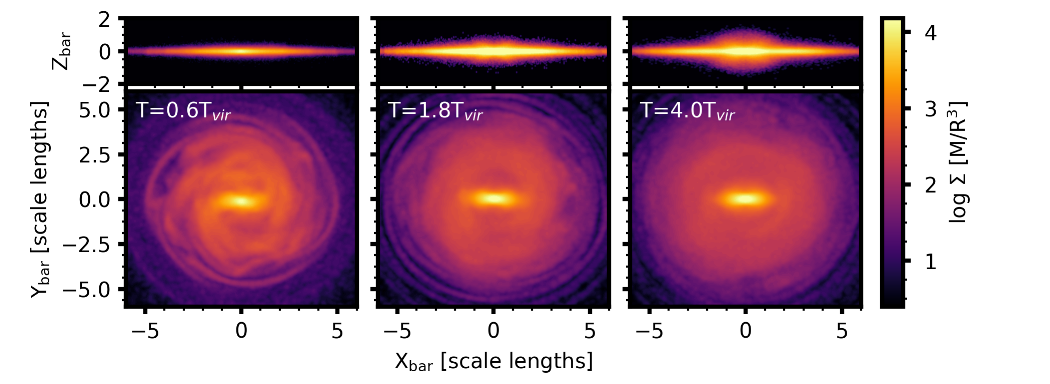} \caption{\label{fig:surfacedensity} Face-on and edge-on snapshots of stellar surface densities at three different snapshots in the cusp simulation. The three snapshots correspond to different different phases of evolution, as discussed in the text: assembly (left panels), growth (middle panels), and steady-state (right panels).} \end{figure*}

\section{Self-Consistent Simulation Interpretation} \label{sec:selfconsistent}

Now we interpret the time-dependent simulations through the lens of the fixed potential orbital structure in Sections~\ref{subsec:cuspsimulation} and \ref{subsec:coresimulation}, and also discuss the limitations of the methodology (Section~\ref{subsec:limitations}). Owing to the caveats discussed elsewhere in this work, we do not expect this to be a complete assessment of the processes at work in the simulation, but rather a dynamical characterisation of the connection between the details of the gravitational potential, the orbital structure, and the relationship between the orbital structure and the resultant evolution. Our study of fixed-potential orbits reveals that orbit families are correlated with distinct bar evolutionary phases. The cusp and core simulations evolve differently in time, but evolve similarly for a similar orbit structure. To see this, we present the evolution of the cusp simulation in Section~\ref{subsec:cuspsimulation}, then the core simulation in Section~\ref{subsec:coresimulation}. We draw comparisons and contrasts between the two in Section~\ref{subsec:selfconsistentsummary}. In each of the first two sections, we describe the overall evolution of the simulation before looking at the underlying orbital structure.

In each simulation, we identify three phases for the bar: (i) assembly, (ii) growth, and (iii) steady-state. The assembly phase begins at roughly the same time for both models. However, the assembly phase is the least likely to be observed in the real universe, owing to the non-cosmological equilibrium initial conditions of the models. For the cusp model, the assembly continues at a slower pace, before transitioning smoothly into the growth phase, then finally reaches a steady state at late times. The core simulation rapidly assembles, then hits a steady-state plateau. At late times, as the bar slowly transfers angular momentum to the halo and the mass distribution rearranges, the bar begins to grow again. In Figure~\ref{fig:surfacedensity} we show example snapshots from the cusp simulation, which correspond to different phases of evolution.

\subsection{Cusp Simulation Evolution} \label{subsec:cuspsimulation}

\begin{figure*} \centering \includegraphics[width=6.5in]{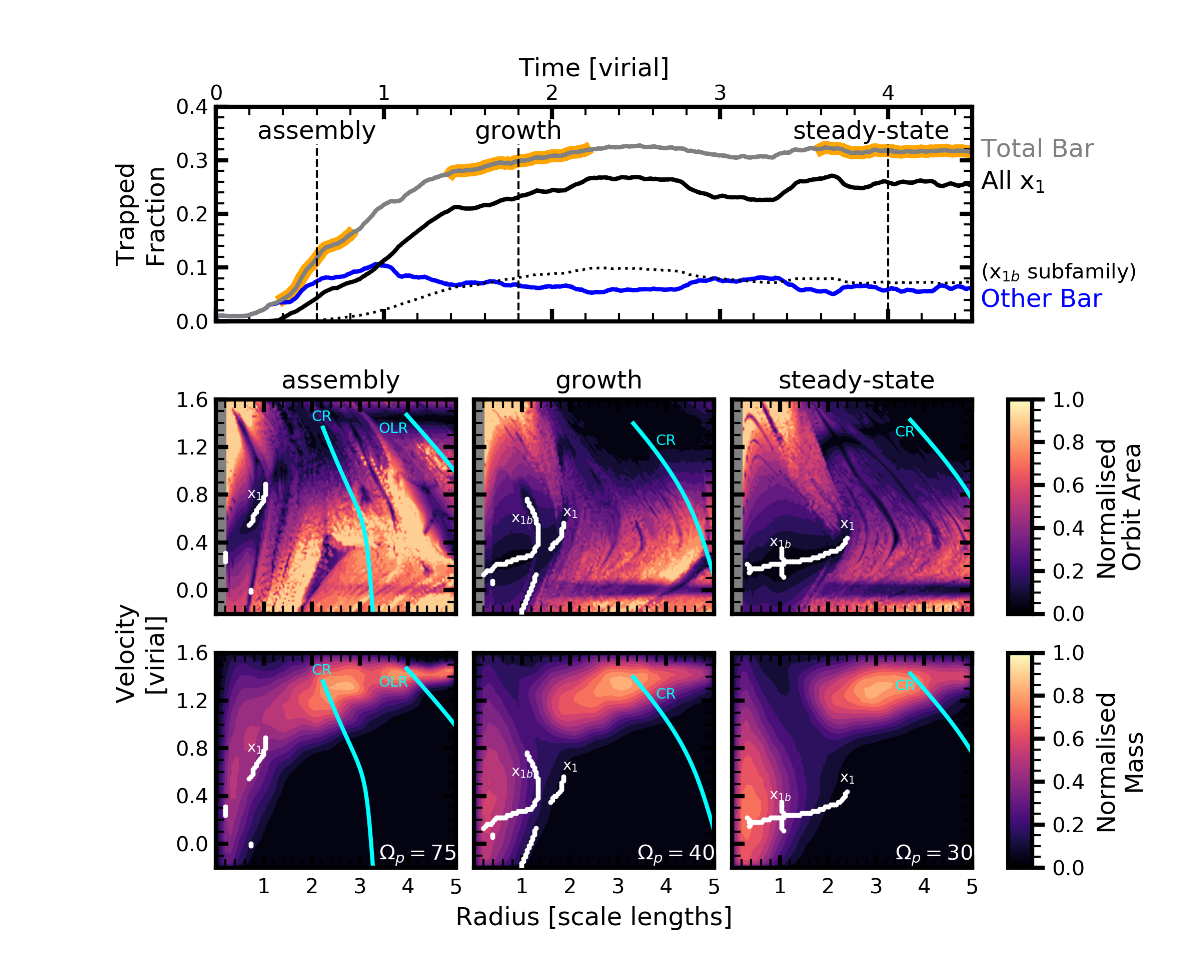} \caption{\label{fig:cusp_timing} Upper panel: Disc trapped fraction versus time for the cusp simulation, in system units. The black curve is the bar-parenting $x_1$ family, the blue curve is a collection of other bar-supporting orbits that are not formally members of the $x_1$ family, and the thin dotted black curve are orbits in the $x_{1b}$ subfamily (a subset of the black curve). The grey curve in the upper panel is the sum of the black and blue lines, which is the total mass trapped in and supporting the bar. Three distinct epochs are highlighted: bar assembly, bar growth, and steady-state. A dashed line indicates the central time for each epoch, when we extract the corresponding potential from the self-consistent cusp simulation and construct an orbit atlas. Middle panels: Computation from the tessellation algorithm at each of the times indicated in the upper panel as a function of $\rturn$ and $\vturn$. The colour indicates $A$, the fraction of the area an orbit fills relative to a circle with the same radius as $\rturn$. Strong commensurabilities are marked and the evolutionary implications are described in the text. White lines correspond to tessellation algorithm-calculated commensurabilities and cyan lines correspond to monopole-calculated commensurabilities. Lower panels: Instantaneous tangential velocity, calculated as $v_T = (x\dot{y} - y\dot{x})/R$ versus instantaneous radius, drawn from the corresponding central times of each epoch (assembly, growth, steady-state) during the simulation. The colour map is normalised relative to the locus of mass for the disc in $\rturn-\vturn$ space. Overlaid on each are the traced commensurabilities as in the middle panels.} \end{figure*}

In Figure~\ref{fig:cusp_timing}, we illustrate the power of using tessellation to find commensurabilities to reveal different mechanisms at work in the cusp simulation. The upper panel shows the trapped fraction versus time for the two populations described in Section~\protect{\ref{subsec:trapping}}. We also include the subfamily of $x_{1b}$ orbits as a dotted black line. The first population to appear are the `other' bar-supporting orbits (blue), whose apoapses show modest coherence with the ends of the bar, but are not part of the $x_1$ family. This is consistent with a standard picture of orbit apoapse precession building the bar. With enough other bar-supporting orbits in place, the $x_1$ family appears (the black line in Figure~\ref{fig:cusp_timing}) at $T=0.4$. The two populations grow in tandem until $T=1$, when the $x_1$ family begins to dominate. Eventually, the rapid assembly of the bar draws to a close at $T=1.4$, and the $x_1$ orbits grow at a slower rate until $T=2.4$, during which time some `other' bar-supporting orbits are converted into $x_1$ orbits. Near $T=3$, the $x_1$ orbits experience an oscillation before the bar stops growing at $T=4$. The dynamics of this oscillation is addressed in a companion work, \citetalias{petersen18c}.

The grey line in the upper panel of Figure~\ref{fig:cusp_timing} depicts the total bar mass, which is the sum of the $x_1$ and `other' bar supporting orbits. We highlight three phases of bar evolution: bar assembly, bar growth, and steady-state. After the assembly and growth phases, the bar is 30 per cent of the entire disc population, in line with estimates for the bar-to-disc ratio in the MW \citep{blandhawthorn16}.

The assembly phase of the bar is marked by nearly equal contributions from the $x_1$ family and `other' bar orbits. With the additional information provided by the closed-orbit map (the lower left panel of Figure~\ref{fig:cusp_timing}), we see that the relatively small mass in the $x_1$ family is expected. CR, while present at $T=0.6$, has begun to migrate outward substantially from its initial position ($r_{\rm CR,~T=0.0}=1.4R_d$, $r_{\rm CR,~T=0.6}=2.0R_d$) but has not yet migrated outward enough to be at a larger radii than the majority of the disc mass distribution. The outer disc, $r>3R_d$, appears nearly unevolved at this time.

The middle panels of Figure~\ref{fig:cusp_timing} are the results of extracting the potential at the times labelled in the upper panel of Figure~\ref{fig:cusp_timing} and integrating the standard initial grid of trajectories described in Section~\protect{\ref{subsubsec:initial_conditions}}, constructing an orbit atlas. As in Figure~\ref{fig:fishbone}, the colour in the middle panels indicates $A$, the fraction of the area an orbit fills relative to a circle with the same radius as $\rturn$, as calculated in Appendix~\ref{appendix:tesselationalgorithm}. The bottom panels are similar to those in Figure~\ref{fig:resonance_positions} but at the labelled times. When present, we mark in the lower panels of Figure~\ref{fig:cusp_timing} the bar-parenting family $x_1$, the bifurcation of the bar-parenting family $x_{1b}$, and the location of corotation orbits, CR, following the procedure in Appendix~\ref{appendix:tesselationalgorithm}. Many other weak higher-order $m$:$n$ resonances are also present, in particular during the bar growth phase, indicative of a rich resonant structure. The differences in the lower panels reveal the mechanisms behind the three distinct phases. During bar assembly, the location of the families in $\rturn-\vturn$ space changes rapidly, resulting in discontinuities and resonances that appear as narrow valleys in $\rturn-\vturn$ space. Prominent $x_1$ and $x_{1b}$ families are present while the bar is in the growth phase. A large density of resonance valleys at the end of the bar ($\rturn\ge2R_d$ and $\vturn\ge1$) and near corotation continue to feed the bar mass as orbits pass through resonances and lose angular momentum. During the growth period, the fraction of $x_{1b}/x_{1}$ orbits increases to a maximum fraction of 40 per cent. When the bar has reached a steady state, the resonant valleys have become more well defined but fewer in number, and the bar orbits have settled into a lower-energy $x_1$ valley as the $x_{1b}$ valley has disappeared. Fewer resonances at the end of the bar and beyond causes the bar growth to stall.

The loss of the $x_{1b}$ family for the $m\le2$ potential (Section~\ref{subsubsec:bcusp}) prompts us to investigate the role of $m>2$ harmonics in the disc, their effect on the $x_{1b}$ population, and the subsequent evolution of the model. To this end, we perform a numerical experiment where we restrict the azimuthal orders in the disc to $m\le2$, after the bar has already formed. {\sc exp} allows for an easy manipulation of different basis functions to investigate the role of individual harmonics on the overall evolution. Specifically, we suppress the $m>2$ terms in the disc of the cusp simulation by applying an error function prefactor. The error function is centred at $T_{\rm off}=1.2 T_{\rm vir}$ with a width of $\delta T_{\rm off}=0.12 T_{\rm vir}$, which corresponds to roughly two bar periods. The $m>2$ coefficients are fully suppressed by $T=1.5$. Construction of closed-orbit maps at $T=1.2 T_{\rm vir}$ and $T=1.4 T_{\rm vir}$, when the prefactor on the $m>2$ terms is 0.5 and 0.0 respectively, conclusively demonstrates that the $x_{1b}$ family is not present when the model is restricted to the dipole and quadrupole terms. Further, the bar rapidly evolves to a new, shorter configuration rather than continuing to grow as in the unmodified cusp simulation. This suggests that the $m=4$ component of the potential is necessary for $x_{1b}$ orbits to exist, and the $x_{1b}$ orbits are part of the main backbone of the long bar, even though the $x_{1l}$ orbits provide the principal observed length.

Identifying orbit families in self-consistent simulations will be necessary for observational comparisons.  For example, \cite{binney91} interpreted observations of gas dynamics toward the centre of the MW to be the result of $x_1$ orbits and the $x_2$ family, eccentric orbits elongated perpendicular to the bar. While the non-bifurcated $x_1$ family becomes more eccentric as one moves inward, as noted by \cite{binney91}, the $x_{1b}$ family remains highly elongated even to the end of the bar, a point that may have observational implications for the MW. Additionally, although we have not discussed the $x_2$ orbits in this paper, our method to compute trapping and the tessellation algorithm are both suitable for the identification and classification of $x_2$ orbits.

\subsection{Core Simulation Evolution} \label{subsec:coresimulation}

\begin{figure*} \centering \includegraphics[width=6.5in]{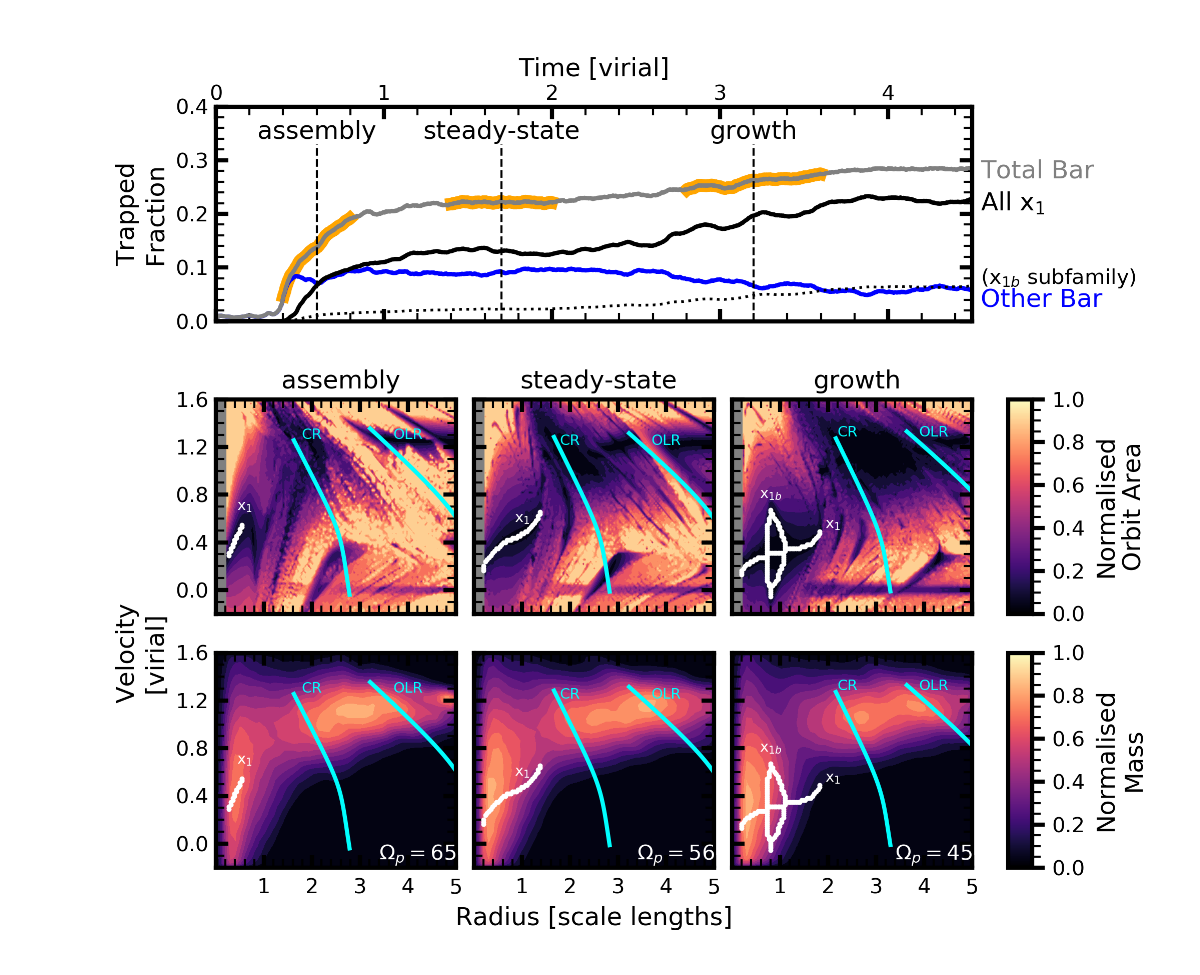} \caption{\label{fig:core_timing} Upper panel: Disc trapped fraction versus time for the core simulation, in system units. The colours are as in Figure~\protect{\ref{fig:cusp_timing}}. Three distinct epochs are highlighted: bar assembly, steady-state, and bar growth. In this simulation, the bar reaches an unstable steady state before growing further. A dashed line indicates the central time for each epoch, when we extract the corresponding potential from the self-consistent cusp simulation and construct an orbit atlas. Middle panels: Computation from the tessellation algorithm at each of the times indicated in the upper panel as a function of $\rturn$ and $\vturn$ in white, with monopole-calculated commensurabilities marked in cyan. Colours and indicators are again as in Figure~\protect{\ref{fig:cusp_timing}}. Lower panels: Instantaneous velocity versus instantaneous radius corresponding to the central time for each evolutionary epoch. The commensurability tracks from the middle panels are overlaid.} \end{figure*}

Some elements of the core simulation are similar to that of the cusp simulation, including the observations of multiple distinct phases of bar evolution. However, a transient steady-state phase precedes the growth phase in this model. As in Figure~\ref{fig:cusp_timing}, the grey line in the upper panel of Figure~\ref{fig:core_timing} measures the total bar mass. After the assembly and growth phases, the bar is 28 per cent of the entire disc population, still in line with observational results. However, for a substantial fraction of time during the simulation, the bar is $<$25 per cent of the total disc population, in contrast to the more massive bar, $\approx 30$ per cent in the cusp simulation.

The bar assembly phase begins at roughly the same time as in the cusp simulation, $T=0.4$. However, the $x_1$ family, shown in black, has already begun to dominate the bar mass fraction by $T=0.6$, when we make the first closed-orbit map. Despite the relative mass equality between $x_1$ orbits and other bar orbits, the closed-orbit map does not show a prominent, contiguous $x_1$ valley from $\rturn=0$ to $\rturn=R_d$. The $x_1$ valley is instead broken by the presence of a second commensurability valley (another $x_1$ bifurcation) at $r=0.8R_d$, which is a result of the rapidly changing potential during the assembly phase.

After $T=0.6$, instead of the `other' bar orbits (blue line) gradually becoming $x_1$ orbits as in the cusp simulation, the two populations remain distinct until $T=2.4$. Having reached a steady state at $T=1.4$, one would be excused for believing the evolution of the system were complete. However one entire time unit later at $T=2.4$, the $x_1$ family begins growing, and captures some of the `other' bar orbits into $x_1$ members. From $2.4<T<3.8$ the bar grows, in much the same manner as the bar growth phase in the cusp simulation. We label this phase bar growth, and calculate the closed-orbit map from the orbit atlas in the bottom right panel of Figure~\ref{fig:core_timing}. During this time period, the fraction of $x_{1b}/x_{1}$ orbits begins increasing, from 15 per cent of the $x_1$ family to a maximum fraction of 30 per cent.

The closed-orbit map for the transient steady-state phase in the lower centre of Figure~\ref{fig:core_timing} resembles the steady-state evolution of the cusp simulation (the rightmost column in Figure~\ref{fig:cusp_timing}), particularly in the number and specific families of observed commensurate valleys. The closed orbit parenting the $x_1$ family and corotation is clear, as well as several other higher-order resonances. By comparison to the lower panel of Figure~\ref{fig:live_quintile}, which is near to panel b in simulation time, we see that the phase space near the resonances is not populated. However, in the bar growth panels, the lower right of Figure~\ref{fig:core_timing}, we see a rich resonant structure, akin to that during the bar growth phase in the cusp simulation. The higher-order resonances have swept outward into the bulk of the phase-space density in the simulation. Additionally, the $x_{1b}$ family has appeared within the bar radius.

\subsection{Limitations of Fixed Potential Analysis} \label{subsec:limitations}

We have presented many orbital snapshots of evolving barred galaxies, however, the orbits themselves evolve with time. While we have learned about the orbital structure at specific times during the secular process, the ongoing temporal evolution is likely to be important in its own right. Future work will attempt a temporal analysis of orbital structure, using the same technique.

A comparison between Figures~\ref{fig:fishbone}, \ref{fig:live_quintile}, and \ref{fig:resonance_positions} suggests that many of the higher order features in Figure~\ref{fig:fishbone} may not exist in the self-consistent simulation owing to system evolution. When the period of the closed orbit is longer than the secular evolution time scale for the evolving barred galaxy, the fixed-potential approximation is inconsistent. Further, we noted subtle but important differences between the fixed potential and self-consistent orbits in Section~\ref{subsec:selfconsistent}, indicating that the fixed potential orbits are missing some dynamics (e.g. transients). The culprit is likely a time-dependent feature that was excluded to make the fixed potential integration stable: (1) odd azimuthal harmonics, (2) harmonic interaction and/or multiple pattern speeds, and (3) a frozen noise spectrum. This will limit the applicability of fixed-potential analyses to both self-consistent simulations and observations.

Inspection of orbits drawn from the self-consistent time series suggest that $x_1$ orbits are robust against the inclusion of odd harmonics in the self-consistent simulation, so it is likely not simply the inclusion of the odd azimuthal harmonics that will create the observed behaviour. Based on a reconstruction of the perturbing $m=1$ disturbance, the peak of the response is at $r<R_d$, and thus unlikely to change the orbital structure in the outer disc more than in the inner disc. To test the role of modal interplay, we have integrated orbits in potential models where the azimuthal series only included the monopole $m=0$ and the quadrupole $m=2$ as a representation of the bar. The quadrupole models disagree with the structure observed in the simulation: for example, $x_{1b}$ orbits do not exist if the bar is represented by a quadrupole only.

All azimuthal harmonics $m>0$ have the same pattern speed imposed. We have checked that this assumption is consistent with the simulation for even harmonic orders. A future investigation will allow for variable pattern speed by azimuthal order. Lastly, it is possible that the choice of any single snapshot may freeze unwanted small-scale noise into the potential. While we believe that the self-consistent field technique will largely smooth out such aphysical fluctuations in the potential, such as small-scale Poisson noise, it is not guaranteed that our implementation of the potential is completely free of aphysical noise on small scales.

Nonetheless, the agreement in identified orbits between the $k$-means classifier of self-consistent orbits (Figure~\ref{fig:example_bar}) and orbits integrated in the corresponding fixed potential (Figure~\ref{fig:orbmap1}) suggests that the sources of uncertainty discussed in this section do not affect our overall interpretation of the key dynamics.

\subsection{Summary} \label{subsec:selfconsistentsummary}

The differences in the evolution between the cusp and core models are easy to describe and difficult to explain. Despite this, one can draw several simple conclusions from the comparison of the evolutionary status in the self-consistent simulation and the features in the closed-orbit maps: \begin{enumerate} \item Bar assembly is a multi-feature event: it produces transients and multiple patterns at small and large scales compared to the bar scale. Attempts to construct the closed-orbit map reveal complex structure in the inner disc ($r<R_d$). \item Despite the apparent differences in evolution of the bar, the bar growth phase in both the cusp and core simulations includes the presence of the $x_{1b}$ family. While the two simulations do not comprise an exhaustive study of parameter space, the similarities in the dynamical mechanism (e.g. the appearance of $x_{1b}$ orbits, fed by 3:$n$ orbits) present an avenue for bar growth.  \item A steady-state phase may either follow (in the case of the cusp simulation) or precede (in the case of the core simulation) the growth phase. \item Despite all that can be gleaned from the closed-orbit map, it must be used with other diagnostics, such as torque analysis \citepalias{petersen18b} and/or harmonic decomposition \citepalias{petersen18c}, to fully interpret simulations. \end{enumerate}

\section{Conclusion} \label{sec:conclusion}

In this paper, we present a new tessellation algorithm for identifying orbit families in arbitrary potentials. The tessellation algorithm generalises fixed potential studies to evolving potentials. We apply the algorithm to two self-consistent simulations (the cusp and the core simulation), from which we select four potentials (at $T=0$ and $T=2$ for each simulation) to learn about the commensurability structure in MW-like models. We fully characterise the orbit structure of the models and completely identify the closed-orbit structure. Our initial conditions were chosen to resemble the MW in disc-to-halo mass ratio, disc scale length to scale height ratio, and general rotation curve shape, but we make little attempt to match the data for the MW beyond scaling the system to match the virial units of the MW. Rather, our aim was a description of dynamical mechanisms that we expect to be phenomena prominent in observed barred galaxies.

Our main finding from the fixed potential analysis is that the resulting `closed-orbit maps' characterise the orbit families admitted by different potential models. The allowed orbit families are sensitive to the shape of the halo profile, allowing for a differentiation between the underlying potential by observing orbits in self-consistent simulations. We also rediscover the $x_{1b}$ family of orbits, referred to as 1/1 orbits in the early literature. We demonstrate that the $x_{1b}$ orbits are harbingers of bar growth. With the tessellation algorithm, we are able to draw connections with self-consistent simulations and analytic works \citep[e.g.]{contopoulos89}. In particular, we interpret previously identified epochs of bar evolution (assembly, growth, and steady state) using closed-orbit maps. The distinctions between different closed-orbit maps, such as the presence or absence of known key orbit families correlates with distinct evolutionary phases of the barred galaxy, and may be used to assess its dynamical state.

We propose a simple new method to interpret IFU data by using a pseudo phase space, the $\rturn-\vturn$ (or $r-v$) plane. For external galaxies, different closed-orbit map models may be compared in the $\rturn-\vturn$ plane to ascertain whether the barred galaxy is in a steady-state or growing phase, based on the location of breaks and features in the $\rturn-\vturn$ plane (cf. Figure~\ref{fig:resonance_positions}).

These methods can connect features of specific orbit families to observed spatial and kinematic signatures in the MW. Observations in the near future (e.g. Gaia, SDSS V) will reveal more about the orbit structure of the inner MW. We predict that if the MW has either a cusped dark matter profile or an old bar, that $x_{1b}$ orbits will be present. If signatures of $x_{1b}$ orbits are observed, these orbits would be an indicator of long-term stability in the bar, as they comprise an extremely stable family in the self-consistent simulations. This work improves upon previous studies of possible orbit structure in a MW-like barred galaxy. In particular, popularly used potentials for the MW, such as {\tt MWPotential2014} from {\tt galpy} \citep{bovy15}, are known simplifications that meet only the most rudimentary requirements for matching the potential of the MW (i.e. a fixed bar pattern speed, lack of transients, missing $m>2$ components of the bar).

Additional applications for this methodology include an extension to other non-separable realistic potentials and studying the rate at which orbits transition between families through coupling to self-consistent simulations. These rates could be connected to simple chemical models to attempt to explain chemically-distinct components of galaxies. In the future, we plan to apply the orbit-volume method in three dimensions and develop potentials for a range of realistic galaxies.

Finally, analyses such as those presented here are just one way of studying the dynamics of barred galaxies. One can also look at the torques as in \citetalias{petersen18b} or harmonics as in \citetalias{petersen18c} to gain further insights.

\section*{Acknowledgements} We thank the anonymous referee who strengthened the results of this paper. We thank Elena D'Onghia for providing simulation snapshots from \citet{donghia19} for a comparison of {\sc exp} and {\sc gadget-3}. MSP thanks Douglas Heggie and Anna Lisa Varri for helpful discussions. This project made use of {\it numpy} \citep{numpy} and {\it matplotlib} \citep{matplotlib}.

\section*{Data Availability}
The data underlying this article will be shared on reasonable request to the corresponding author.

\bibliography{PetersenMS}

\appendix

\section{Orbit Classification} \label{appendix:orbitclassification}

For an orbit conserving $L_z$ and $E$, the maximum radius for a given orbit (apoapse) is set by the orbital frequencies. As we are most interested in the classification of orbits related to the bar, we evaluate the apoapse locations relative to the minimum of the bar potential, i.e. the bar position angle.  We use this distance to classify orbits into different families with a clustering algorithm. We use $k$-means \citep{lloyd82} to partition apoapses for a given orbit into families that show similar morphologies. We then connect to known classical orbit families. Briefly, the $k$-means algorithm iteratively separates a collection of points into $k$ clusters by minimising the sum of the distance between each point and the centre of a determined cluster. The distance metric used in this work is standard Cartesian space.

Since our focus is on the bar, we restrict our analysis here to $k=2$ clusters. One may naturally understand this choice as an attempt to find orbits that exhibit apoapses clustering at either end of the bar. An $x_1$ orbit, such as that shown in panel a of Figure~\ref{fig:example_bar}, has apoapses that are strongly clustered in two points in space, i.e. at either end of the bar. An orbit that is bar-supporting, but distant in phase-space from the parent $x_1$ orbit, will show apoapses that are loosely clustered around the end of the bar\footnote{Some orbits may show four-fold symmetry and naturally have a minimum distance from the cluster centre when $k=4$; however, the cluster centres are degenerate with a $k=2$ cluster classification, where the $k=2$ cluster centres are an average of two of the $k=4$ cluster centres. The two averaged $k=2$ cluster centres are centred on the end of the bar when a four-fold symmetry orbit is trapped (e.g. panel c of Figure~\ref{fig:example_bar}).}.

We consider each orbit individually and undertake the follow process for each orbit. We prepare the orbit for $k$-means analysis as follows: \begin{enumerate} \item Extract $x,y$, and $z$ time series for a given orbit. \item Determine the apoapses. \item Transform the $x,y$ positions of the apoapses into a frame where the bar position angle is aligned with the $x$ axis. \item For each apoapse: determine the 19 other nearest apoapses in time. \item For each apoapse: partition these 20 apoapses into $k=2$ clusters based on their $x-y$ positions. \item Save the cluster data for each orbit at each apoapse. \end{enumerate} The cluster data includes: (1) the set of apoapses partitioned into each of the $k$ clusters, and (2) the location of the spatial centre (equivalently listed in $(R_{\rm cluster},\phi_{\rm cluster})$ or $(x_{\rm cluster},y_{\rm cluster}$) for each cluster. Using these data, bar membership is determined by applying criteria to the following four metrics: \begin{enumerate} \item $\tparaL\equiv{\rm max}\left(\langle \phi_{\rm bar}\rangle_N\right)_k$, the trapping metric from \citetalias{petersen16a} that assesses the average angular separation in radians from the bar axis, $\phi_{\rm bar}$, for $N$ apoapses in $k$ clusters. $\tparaL$ is the maximum over the $k=2$ mean angular separations for each cluster\footnote{$\tparaL=0$ for four-fold symmetric orbits.}. $N$ is a parameter set based on the dynamical time of the bar, which we set to $N=20$ for all analyses in this work. \item $\left \langle \sigma_{R_{\rm cluster}}\right\rangle_k$, the standard deviation in radius for all apoapses in a cluster, averaged over $k$ clusters. A larger value of $\left \langle \sigma_{R_{\rm cluster}}\right\rangle_k$ relative to $\left \langle R_{\rm cluster}\right\rangle_k$ implies no trapping. A threshold on this ratio effectively removes false positive detections. \item $\left \langle \sigma_{\phi_{\rm cluster}}\right\rangle_k$, the standard deviation in position angle for all apoapses in a cluster, averaged over $k$ clusters. Variation in this quantity is the product of both uncertainty in the bar angle as well as being possibly indicative of a family that would be better fit by an increase in the number of clusters $k$.  \item $\Omega_r$, the instantaneous radial frequency, computed as the finite difference in time between the central apoapse and the next nearest apoapse in time. This quantity is used to calculate orbits with $\Omega_r$ above the Nyquist frequency for time sampling input.  \end{enumerate}

For an $x_1$ orbit, the $k$-means classifier is well-behaved: in the limit where the orbit is not evolving, the apoapses are perfectly confined to two locations in $(x_{\rm bar}, y_{\rm bar})$ space. That is, the clusters will be compact: Thus, $\tparaL\to0$, $\left \langle \sigma_{R_{\rm cluster}}\right\rangle_k\to0$, and $\left \langle \sigma_{\phi_{\rm cluster}}\right\rangle_k\to0$. Therefore, we use these parameters as the primary metrics to construct the classification scheme.

Deviations from zero for the three quantities may result from orbit family switching -- either because the potential is changing or because the orbit is near a heteroclinic trajectory -- or apoapse position uncertainty (owing to discrete time sampling). The deviations require some tolerance to be set on each quantity. We select classification thresholds by constructing diagrams in $(\tparaL, \left\langle \sigma_{R_{\rm cluster}}\right\rangle_k, \left \langle \sigma_{\phi_{\rm cluster}}\right\rangle_k)$ space and looking for features. In practice, at late times, the features are obvious as the orbit evolution has slowed significantly.

We find false positive and negative rates of approximately one per cent. We determined these rates empirically by varying the size of the apoapse window and examining the orbits whose classifications change. We also examine orbits by visual inspection to check for contaminants. We find that the mass determination in the trapped-orbit population is approximately one percent, which is more than sufficient for our dynamical analysis.

The tolerance values for each family are presented in Table~\ref{tab:families}. Choosing the thresholds for family classification is largely model-independent: in the context of this work, we find that the metrics chosen for the cusp simulation correctly classify the orbits in the core simulation as well. The largest difference between the two simulations in terms of classification is in the `other' bar-supporting orbit family, which has some dependence on upon the geometry of the bar (e.g. a more rectangular bar has more orbits with nonzero $\left \langle \sigma_{\phi_{\rm cluster}}\right\rangle_k$). In addition, time evolution can affect the time resolution; determinations are nosier in models with rapidly changing pattern speeds. To calibrate the thresholds for trapped orbits given the five metrics, we tabulate the quantities listed above for all orbits at some late time, when secular evolution is relatively slow. Significant progress on tuning the thresholds may be made using theoretical considerations: orbits that are part of the bar will show small values of $\tparaL$, and orbits that are consistent members of a single family will show small values of $\left \langle \sigma_{R_{\rm cluster}}\right\rangle_k$. We empirically find the best discriminator between different orbit families to be $\left \langle \sigma_{\phi_{\rm cluster}}\right\rangle_k$. We assume that orbits with $\Omega_r>\Omega_p$ are not a part of the bar owing to classification uncertainty as a result of orbit sampling. Once we complete the process for verifying the chosen criteria, we proceed with a full analysis of the simulation.

We identify four major improvements over \citetalias{petersen16a}: \begin{enumerate} \item Implementation of the `k-means++' technique of \cite{arthur07} when the standard (Lloyd's) k-means technique \citep{lloyd82} fails (approximately 0.6 per cent of orbits in the cusp model). \item Use the closest $N$ apoapses in time to the indexed time, rather than enforcing $\frac{N}{2}$ apoapses on either side of the target time. \item Set a threshold, $T_{\rm thresh}$, that is some multiple of the bar period $T_{\rm bar}$ in which the $N$ apoapses must reside. This guards against choosing unrelated apoapses. The threshold is exceeded for approximately 15 per cent of the fiducial model orbits, at which point the orbit is not analysed at that timestep. \item The inclusion of $\left \langle \sigma_{\phi_{\rm aps}}\right\rangle_k$ allows for a subdivision into 2:1, 4:1, and higher order families even when using $k=2$. \end{enumerate}

Table~\ref{tab:families} lists the empirically-determined classification criteria for two families of orbits in our cusp and core simulations. Table~\ref{tab:families} also lists an `Undetermined' classifier for orbits with $\Omega_r$ above the Nyquist frequency of the time-series sampling. These are nearly always orbits that are close to the centre. Additionally, while $x_{2,3,4}$ orbits exist in small quantities in our models, these orbits play little if any role in the dynamics described in this work, and so we do not focus on their classification here.

\begin{table} \label{tab:families} \begin{tabular}{lcccc} \hline Family & $\langle \phi_{\rm bar} \rangle_{20}$& $\Omega_r$ & $\left \langle \sigma_{r_{\rm aps}}\right\rangle_k$ & $\left \langle \sigma_{\phi_{\rm aps}}\right\rangle_k$ \\ \hline $x_{1}$ & $\left[0,\frac{\pi}{6}\right]$ & $<\frac{1}{2\delta t}$ & $\left[0,0.1a\right]$ & $\left[0, \frac{\pi}{16}\right]$ \\ Other Bar &$\left[0,\frac{\pi}{6}\right]$ & $<\frac{1}{2\delta t}$ & $\left[0,0.1a\right]$ & $\left[\frac{\pi}{16}, \frac{\pi}{8}\right]$ \\ Undetermined & - & $>\frac{1}{2\delta t}$ & - & - \\ \hline \end{tabular} \caption{Membership definitions for being classified into families. `-' indicates that no constraint was placed on the parameter. $\langle \phi_{\rm bar} \rangle_{20}$ is the average angle of the cluster centre relative to the bar major axis over 20 radial periods. $\Omega_r$ is the radial frequency of the orbit. $\left \langle \sigma_{r_{\rm aps}}\right\rangle_k$ is the standard deviation in radius of apoapses from the radius of the cluster centre to which the apoapses belong. $\left \langle \sigma_{\phi_{\rm aps}}\right\rangle_k$ is the standard deviation in angle of apisides from the angle of the cluster centre to which the apoapses belong. } \end{table}

We also would like to identify the bifurcated members of the $x_1$ family, the $x_{1b}$ orbits, which we refer to as a subfamily. We employ a secondary classification scheme on orbits that we determine to be part of the larger $x_1$ family. Unfortunately, the metrics for the apoapses of $x_{1b}$ and $x_{1s,l}$ orbits are indistinguishable. That is, both $x_1$ and $x_{1b}$ orbits will satisfy the criteria set forth for $x_1$ orbits in Table~\ref{tab:families}.  However, a clear morphological difference in the bifurcated $x_{1b}$ orbits and the short- or long-period $x_1$ orbits is the presence of varying numbers of $x_{\rm bar}$ and $y_{\rm bar}$ local maxima. Equivalently, these are points where the velocity in one dimension relative to the bar goes to zero. Therefore, a classification scheme that separately identifies local maxima in the time-series of $x_{\rm bar}$ and $y_{\rm bar}$ distinguishes between the subfamilies. Such a classification scheme is computationally expensive, requiring tracking the entire time-series for a given orbit after it has been identified as an $x_1$ orbit. As shown in this work, $x_1$ orbits may be 30 per cent of all disc orbits in the simulation.

Despite this uncertainty, we make estimates of membership in the $x_{1b}$ family from the $x_{\rm bar}$ and $y_{\rm bar}$ frequencies, $\Omega_{x_{\rm bar}}$ and $\Omega_{y_{\rm bar}}$, as determined by the local maxima of the $x_{\rm bar}$ and $y_{\rm bar}$ time series. The $x_{1b}$ orbits trace two morphologies: infinity symbol-like orbits, and smile- or frown-like orbits. Infinity symbol orbits have $\Omega_{x_{\rm bar}}/\Omega_{y_{\rm bar}}=1.5$. Smile and frown orbits have $\Omega_{x_{\rm bar}}/\Omega_{y_{\rm bar}}=2$ because the strongest smiles and frowns counter-rotate in the bar frame. The subclassification into $x_{1b}$ orbits benefits from this distinction with the standard $x_1$ orbits, which have $\Omega_{x_{\rm bar}}/\Omega_{y_{\rm bar}}=1$ or $\Omega_{x_{\rm bar}}/\Omega_{y_{\rm bar}}=3$ (in the case of $x_1$ orbits with so-called `ears', see Figure~\ref{fig:example_orbits}). Far away from the closed orbit, the classification between $x_1$ and $x_{1b}$ becomes subjective. Therefore, we offer only a coarse estimate of the membership, assuming that orbits with $1.5\le\Omega_{x_{\rm bar}}/\Omega_{y_{\rm bar}}\le2$ are $x_{1b}$ orbits, which we can then classify into infinity or smile/frown orbits by the presence or absence of counter-rotation in the rotating frame. The broad classification by frequency is necessary as orbits do not spend large fractions of time as closed-orbit members of the subfamilies, with small integer combinations of $\Omega_{x_{\rm bar}}$ and $\Omega_{y_{\rm bar}}$, but rather exhibit modest resemblance to the parent orbit as secular evolution proceeds.

We estimate the membership in the bifurcated families in this work. Owing to the uncertainty for any given orbit at a particular time, we consider orbits only over large windows of time during our analysis, reducing the uncertainty, but limiting the time resolution of our estimates for the fraction of $x_{1b}$ orbits to $\delta T_{\rm x_{1b}~sampling}=0.1T_{\rm vir}$. This coarse time resolution is sufficient to track global trends in the $x_{1b}$ subfamily relative to the overall $x_1$ membership.

Lastly, the sign of the maxima can also help determine the preferred orientation of bifurcated orbits, which are asymmetric with respect to either the bar major axis (in the case of `symmetric' $x_{1b}$ orbits) or the bar minor axis (in the case of `asymmetric' $x_{1b}$ orbits). That is, we may determine whether the crossing point in the infinity orbits is preferentially located toward one end of the bar, or whether the counter-rotating portion of the smile/frown orbits, such as the example in panel `b' of Figure~\ref{fig:orbmap1}, is toward one direction along the bar minor axis. Such an asymmetry of crossing points or counter-rotating directions is responsible for the $m=1$ mode, discussed in \citetalias{petersen18c}.

\section{The Tessellation Algorithm} \label{appendix:tesselationalgorithm}

\subsection{Orbit Area Measurement}

Given a series of samples at discrete times for an orbit, we wish to approximate the area that an orbit would sample in the theoretical limit where the time interval between points in the series goes to zero, $\delta t\to0$ and the total time integration is very long, $T\to\infty$. To measure the area of an annulus or volume of a sphere that a discrete set of orbit time samples would eventually fill as $T\to\infty$, so we require a tessellation technique that transforms a discrete time series of points into an integrable area, $\delta t\to0$.  One such computational technique is Delaunay triangulation. We construct a procedure that uses Delaunay triangulation (DT), taking as input a set of points and returning a single value that is the (normalised) computed orbit area from the sum of individual tessellated triangles\footnote{An equivalent procedure may be followed to generalise the orbit measurement to a volume, i.e. by using all three dimensions for the orbit. In such a procedure, triangles become tetrahedrons from which a volume can be computed.}.

The steps to calculate the area of a given orbit from the time series of discrete points are as follows: \begin{enumerate} \item Integrate the orbit in a given rotating potential using discrete timestep $dt$ as in Section~\ref{subsubsec:integration} to obtain a set of two-dimensional points in the bar frame.  \item Compute the triangulation of the transformed points by applying DT to the $(x,y)$ orbit points. \item Prune the triangulation by eliminating triangles with axis ratios above some threshold. \item Compute the area of each remaining triangle and sum to obtain $A$, the area of the orbit. \item (optional) Normalise the area of the orbit by the area of a circle with radius $r_{\rm max}$, the maximum distance from the inertial centre in the time series.  \end{enumerate}

We use the computational geometry library CGAL \citep{cgal} to perform the DT. From the input of a time series of vertices comes a list of triangles with length-ordered sides \(a\ge b\ge c\). We trim to remove the 'webbing' that would connect parts of the orbit that are at unrelated orbital phases by ignoring triangles with extreme axis ratios, which we choose to be $\frac{a}{c}>10$. Adjusting this threshold to 5 or 15 does not produce qualitatively different results.

\begin{figure} \centering \includegraphics[width=2.8in]{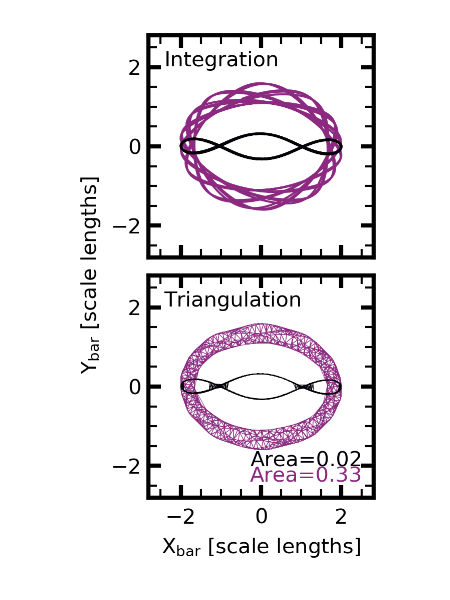} \caption{\label{fig:example_orbits} Two orbits in the barred cusp potential, Potential II. The black orbit starts from $(\rapo,~\vapo)=(0.02,0.45)$, and the purple orbit starts from $(\rapo,\vapo)=(0.02,1.05)$. The upper panel shows the integration of the orbits in the potential for $\Delta T=0.64$, with the orbits presented in a frame co-rotating with the bar. The lower panel shows the same orbits, with the fraction of a circle that the orbit fills in (area) computed using the tessellation algorithm. Some residual triangles not successfully trimmed by the simplex rules are seen in the black orbit, limiting the absolute precision of the technique.} \end{figure}

Examples of the technique are shown in Figure~\ref{fig:example_orbits}. The upper panel shows the integration of two orbits over the entire time window, $T=0.64$, which is 2000 steps at $h=3.2\times10^{-5}$. Both orbits are shown in the frame co-rotating with the bar. While the black orbit has sampled the entire phase-space trajectory, the purple orbit has not. It is clear that the purple orbit will fill an entire torus in physical space given enough time. In the lower panel, we show the triangulation for each orbit. The black orbit, which is closed, features vanishingly small triangles, while the purple orbit, which previously only sampled a fraction of the torus, is now filled with triangles. We can now evaluate the area in physical space that each orbit occupies.

As described in the optional final step above, to compare the area of orbits with hugely different energies, we normalise by the area of a circle with a radius equal to the maximum radius of the orbit over the course of integration: $A_{\rm norm} = \sum^k T_k/(\pi \rapo^2)$. This yields values $A_{\rm norm} \in (0,1]$. As we always opt to normalise the area, we eliminate the subscript and simply refer to the normalised area as $A$ throughout this work. In the bottom panel of Figure~\ref{fig:example_orbits}, we list the area for both orbits; the closed black orbit has an area $A=0.01$, demonstrating the uncertainty owing to the triangulation (a closed orbit has zero area, $A=0$). The purple orbit has an area $A=0.33$.

\subsection{Orbital Skeleton Tracing}

As shown in Figure~\ref{fig:example_orbits}, a commensurate orbit will occupy a smaller area in physical space than a non-commensurate orbit. We exploit this to find commensurate orbits in the potential. The locations of true commensurate orbits occupy a vanishingly small volume in phase space, suggesting that tracing valleys in orbit area will follow commensurate orbit tracks. The procedure we use to trace commensurabilities in the $\rturn-\vturn$ plane is as follows: \begin{enumerate} \item Identify all orbits below a certain threshold in normalised area. We use $A\le0.02$, which balances the finite measurement accuracy from the triangulation while still excluding non-commensurate orbits.  \item Connect contiguous areas using a standard marching-squares algorithm that checks adjacent orbits in the $\rturn-\vturn$ grid to determine which adjacent orbits meet the threshold criteria and subsequently connect the points, which we call the threshold map.  \item Perform valley-finding on the threshold map using the algorithm of \cite{steger98}. The algorithm calculates the Hessian matrix of the threshold map by convolving the threshold map with derivatives of a Gaussian smoothing kernel and then determining the vanishing point of the gradient, i.e. a valley. \item Inspect individual orbits in the atlas near the commensurability for connection to known families of orbits and frequency ratios. \end{enumerate}

\subsection{Monopole-calculated Commensurabilities}

One may numerically compute the location of resonances in an axisymmetric potential by determining the table of frequencies for each energy $E$ and $X\equiv L_{z, {\rm orbit}}/L_{z, {\rm circular}}$ in a grid and solving equation~(\ref{eq:resonances}) for a given combination of $(m,l_1,l_2,l_3)$. For our expansion of the gravitational potential fields in harmonic orders, we create a spherically-symmetric monopole from the mass distribution. For orbits outside of the bar radius, at evolutionary stages after the bar has formed, the monopole is a reasonable approximation for the potential in the plane. For our analysis restricted to the $(x,y)$ plane, $l_3=0$.

We calculate the $(E,X)$ locations of CR $(m,l_1,l_2) = (2,0,2)$ and OLR $(m,l_1,l_2) = (2,1,2)$ using the monopole approximation. That is, we create a spherically symmetric potential and directly solve for the frequencies that correspond to the desired commensurability. To place the $(E,X)$-calculated locations of resonances on the $(\rturn,\vturn)$-based figures in this work, we compute the transformation between $(\rturn,\vturn)$ and $(E,X)$. In the axisymmetric case, the mapping is monotonic.  The location of ILR, $(m,l_1,l_2) = (2,-1,2)$, is not possible to approximate using this method owing to the strongly non-axisymmetric potential at those radii. In this case, the tessellation algorithm and orbital skeleton tracing are preferred.

\end{document}